\documentstyle[galley,graphicx,epsf]{mn2e}

\begin{document}

\title {Searching for the metal-weak thick disc in the solar neighbourhood}

\author[Reddy \& Lambert]
        {Bacham E. Reddy$^{1}$, David L. Lambert$^{2}$\\
$^{1}$  Indian Institute of Astrophysics, Bangalore 560034, India\\
$^{2}$  The W.J. McDonald Observatory, University of Texas at Austin, Austin, Texas 78712, USA}

\maketitle

\label{firstpage}

\begin{abstract}

An abundance analysis is presented of 60 metal-poor stars drawn from 
catalogues of nearby stars provided by Ariyanto et al. (2005)
and Schuster et al. (2006). In an attempt to isolate a sample of
metal-weak thick disc stars, we applied the kinematic criteria
$V_{\rm rot} \geq 100$ km s$^{-1}$, $|U_{LSR}| \leq 140$ km s$^{-1}$, and  
$|W_{LSR}| \leq 100$ km s$^{-1}$. Fourteen  stars  satisfying these
criteria and having [Fe/H] $\leq -1.0$ are included in the sample of 60
stars. Eight of the 14 have [Fe/H] $\geq -1.3$ and may be simply thick disc stars
of slightly lower than average [Fe/H]. The other six have [Fe/H] from $-1.3$ to
$-2.3$ and are either metal-weak thick disc stars or halo stars with
kinematics mimicking those of the thick disc. 
The sample of 60 stars is completed by eight thick disc stars, 20 stars of
a hybrid nature (halo or thick disc stars), and 18 stars with kinematics
distinctive of the halo.
 
\end{abstract}


\section{Introduction}

Stars in the the Sun's immediate neighbourhood belong in the main to the Galactic
disc with a sprinkling of halo stars. The majority of the disc
stars are from the thin disc with a  minority from the thick disc.
Introduction of the concept of the thick disc is broadly  attributed  to
Gilmore \& Reid (1983) who from star counts 
found that a double-exponential offered a fit to the
space density perpendicular to the Galactic plane toward
the south Galactic pole. The dominant
component with a scale height of 300 pc was the familiar disc, now
referred to as the thin disc. The second component with a scale height
of 1350 pc was dubbed the thick disc. At the Galactic plane, the
thick disc stars comprise no more than a few per cent of the 
thin disc population.
 Obviously, the fractional population represented by the
thick disc increases with height above the plane.

After its introduction, the thick disc was a controversial innovation for
quite some years but,
today, many properties of the thick disc in the solar neighbourhood
are rather well determined.
Many studies have found thick disc stars to be old with ages in the
range of 8-13 Gyr (Fuhrmann 1998; Reddy et al. 2006) but other
studies have proposed ages as young as 2 Gyr for some thick disc stars
(Bensby et al. 2007).
Their velocity dispersion perpendicular to the Galactic
plane is about 40 km s$^{-1}$, a value to be compared with 
about 20 km s$^{-1}$ for the old thin disc, and 90 km s$^{-1}$ for the
halo. Thick disc stars lag behind thin disc stars in rotation about the
centre of the Galaxy by about 50 km s$^{-1}$. The mean metallicity of
the thick disc is about  [Fe/H]  $= -0.6$ with most
stars falling in the interval [Fe/H] of $-0.3$ to $-1.0$. Differences
in relative abundances [X/Fe]
 between thin and thick disc stars are now well established
over this [Fe/H] range (Bensby et al. 2005; Reddy, Lambert, \& Allende
Prieto 2006). 

Remaining significant controversies surround the upper and lower bounds to the
[Fe/H] distribution function for thick disc stars. In this
paper, we are concerned with the lower bound for thick disc stars in
the solar neighbourhood. Stars of the thick
disc with [Fe/H] less than about $-1$ are referred to as `metal-weak
thick disc' stars (here, metal-weak thick disk stars).
The metal-weak thick disk has now been the subject of many investigations.
But,
although many studies of putative metal-weak thick disk stars exist, the results as
far as their composition are concerned  have been limited to a
measurement of metallicity, usually [Fe/H]. In broad terms, two
kinds of samples have been discussed: stars in 
the Sun's immediate neighbourhood
and  giant stars beyond the solar neighbourhood. The significant
advantage provided by local stars is that the sample
has  well determined kinematics. 
The  advantage of a sample composed of stars at greater distances from the
Galactic plane is that
the relative fraction of thick to thin disc stars will be higher than locally but
the obvious disadvantage is, in general, that knowledge of stellar
kinematics is compromised by the lack of accurate distance and
the small amplitude of the proper motions. In addition, halo stars provide an increasingly
severe contamination of stellar samples as distance from the Galactic plane increases.  

The persisting  controversy about the existence of
metal-weak thick disk stars
 may be traceable to reconsideration of the
metallicities obtained  by Morrison
et al. (1990) from DDO photometry for their sample of giants with disclike
kinematics.
 Ryan \& Lambert (1995) undertook a high-resolution
spectroscopic analysis of a subset of Morrison et al.'s
stars and showed that the most of the stars identified as
belonging to the metal-weak thick disk have metallicities [Fe/H] $>$ $-1$ and
are thus `normal' thick disc stars. Twarog \& Anthony-Twarog
(1994, 1996) provided a recalibration of the DDO photometry that
also weakened Morrison et al.'s evidence for existence of
metal-weak thick disk stars.
These critiques of the pioneering paper on metal-weak thick disk stars have
percolated through the literature. 
Beers et al. (2002),
ardent advocates for metal-weak thick disk stars as an important
component of Galactic structure, wrote 
that  `acceptance of their [metal-weak thick disk stars]
presence has been cast in doubt because of incorrectly assigned
metallicities'. Oddly, major studies of  candidate
metal-weak thick disk stars have rarely obtained the metallicity and never elemental
abundances from high-resolution high-signal-to-noise ratio 
spectra.

In this paper, we  search for metal-weak thick disk stars among two recent  catalogues of
local main sequence stars with reliable kinematics and metallicities.
A comprehensive abundance analysis is undertaken for
metal-weak thick disk candidates and halo stars in order to search for definitive
differences in composition between candidate metal-weak thick disk stars and
halo stars of similar metallicity.

\section{Two Samples of Local Metal-poor Star}

The catalogues searched for  possible metal-weak thick disk stars 
are provided by Arifyanto et al. (2005)
and Schuster et al. (2006). 
 These two catalogues give
space velocities $U$, $V$, and $W$ along with the input astrometry
(proper motion, parallax and radial velocity) and a measure of metallicity.
 To maintain uniformity
with our earlier surveys (Reddy et al. 2003, 2006),
we have recomputed space velocities $U$, $V$, and $W$ and expressed them
 relative to the local standard of
rest (LSR), i.e., $U_{\rm LSR}$, $V_{\rm LSR}$, 
 and $W_{\rm LSR}$, 
 using the 
astrometry given in the respective
catalogues and the  solar motion of $(U,V,W) = (10.0, 5.3, 7.2)$ km s$^{-1}$ 
(Dehnen \& Binney 1998)
 where the positive
 values of $U$, $V$, and $W$ are towards Galactic center, in the direction
of Galactic
rotation, and towards the Galactic north pole, respectively.

The differences between our computed  $(U,V,W)$ velocities
and the values given by Arifyanto et al. and  by Schuster et al.
are very small ($\leq$ 0.5 km s$^{-1}$) and quite unimportant in a
selection of metal-weak thick disk stars.
The difference between  $V_{LSR}$ computed
here and the $V_{LSR}$ given by Schuster et. al is $-$9.4$\pm$0.1 km s$^{-1}$   
 due 
to the difference in the correction  of velocities to the LSR: Schuster et al. used
a solar motion of $(U,V,W) = (10.0, 14.9, 7.7)$ km s$^{-1}$, a difference of
9.6 km s$^{-1}$ in $V$.

The Galactic orbital parameters  $R_{\rm min}$ and $R_{\rm max}$, 
the peri-and apogalactic distances, respectively, 
$Z_{\rm max}$ (maximum vertical distance from
the Galactic plane), and the orbital eccentricity $e$ were computed using a Galactic
potential integrator kindly provided by D. Lin (see Reddy et al. 2003). 
Orbital parameters were computed
assuming the Sun's distance to be 8.5 kpc from the Galactic center. The computed orbital
parameters $R_{\rm m}$, $e$, $Z_{\rm max}$ for the sample stars are given in Tables~1 $\&$ 2.
The value $R_{\rm m}$ is the mean of $R_{\rm max}$ and $R_{\rm min}$.

The uncertainties in the derived $U, V, W$ values are due mainly to  
uncertainties in the parallaxes. 
Uncertainties of the proper motions, in general, do not exceed 1$\%$. 
Most of the  selected stars have parallax errors of less than 15$\%$. 
Four stars 
have parallax errors between 20-25$\%$.
 Errors in $U,V,W$ are estimated by computing
the quadratic sum of errors due to errors in parallaxes, 
proper motions, and radial velocities.

From the combined catalogue (Arifyanto et al. plus Schuster et al.), we
initially selected 60 stars for abundance analysis (see below).
 Errors in $U,V,W$ 
are less than 20 km s$^{-1}$  for two thirds of the  sample.
For about 20 stars errors are in between 20 - 50 km $^{-1}$.
Large errors are, mostly, found for halo stars which have large velocities. 
Out of 60 stars in our study, 33 stars
have Hipparcos astrometry (Perryman et al. 1997), which we have adopted 
except for three stars: G060-026, G090-003, and G016-013. 
(Arifyanto et al. adopted Hipparcos data
for all the Hipparcos stars except for those which have parallaxes less than 5 mas)
For the latter 
two, we used photometric parallaxes from Arifyanto et al. catalogue, and for G060-026 astrometry is taken
from the Schuster catalogue. The remaining 30 stars are compared with the new astrometry
deduced from recalibration of the Hipparcos data (van Leeuwen 2007).
The difference in
proper motions between old and recalibrated data
is less than 1$\%$ for most stars and for 4 stars it is
in between 1-5$\%$.
The difference in parallaxes between the two datasets  is, in general,
less than 10$\%$. For
seven stars, it is as high as 20$\%$.
The differences in astrometry result in differences of
about 5 - 10 km s$^{-1}$ in at least one  velocity component.

\begin{figure*}
\includegraphics[width=15cm]{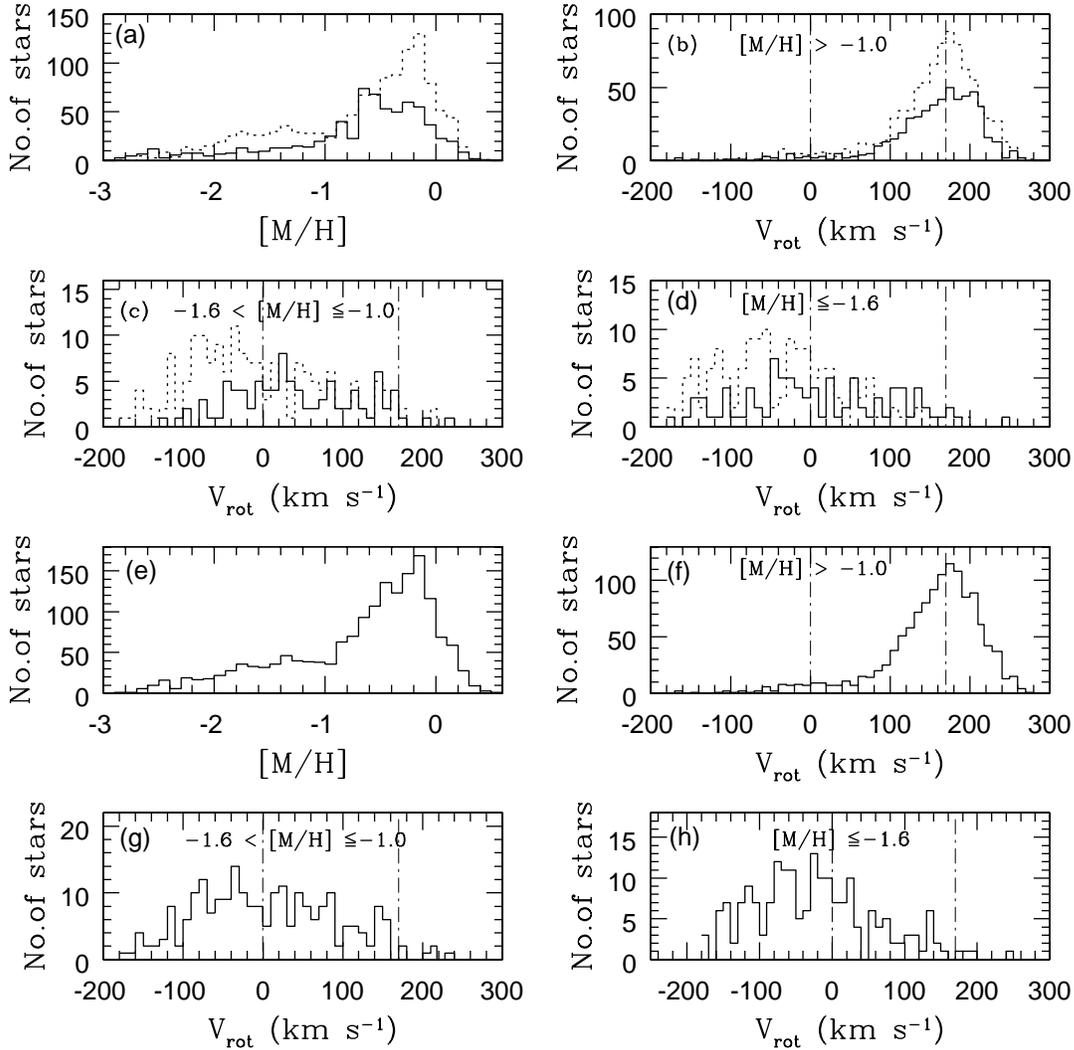}
%\epsffile{arif_allvrot.ps}
%\epsfxsize=18truecm
\caption{Top four panels show distribution functions 
for the  740 stars from Arifyanto et al.'s catalogue (solid line) and
for the 1211 non-binary stars from Schuster et al.'s catalogue (dotted line).
In the panel (a), metallicity distribution is shown for both the catalogues
and in the panels (b) to (d) distribution functions of V$_{\rm rot}$ for three metallicity
ranges are shown. 
Similarly, in the botten four panels (e) to (h), 
distribution functions for the combined sample of 1713 stars are shown.
The two vertical lines are drawn at V$_{\rm rot}$ = 170 km s$^{-1}$ and 0 km s$^{-1}$.} 
\end{figure*}

\begin{figure}
\includegraphics[width=9cm]{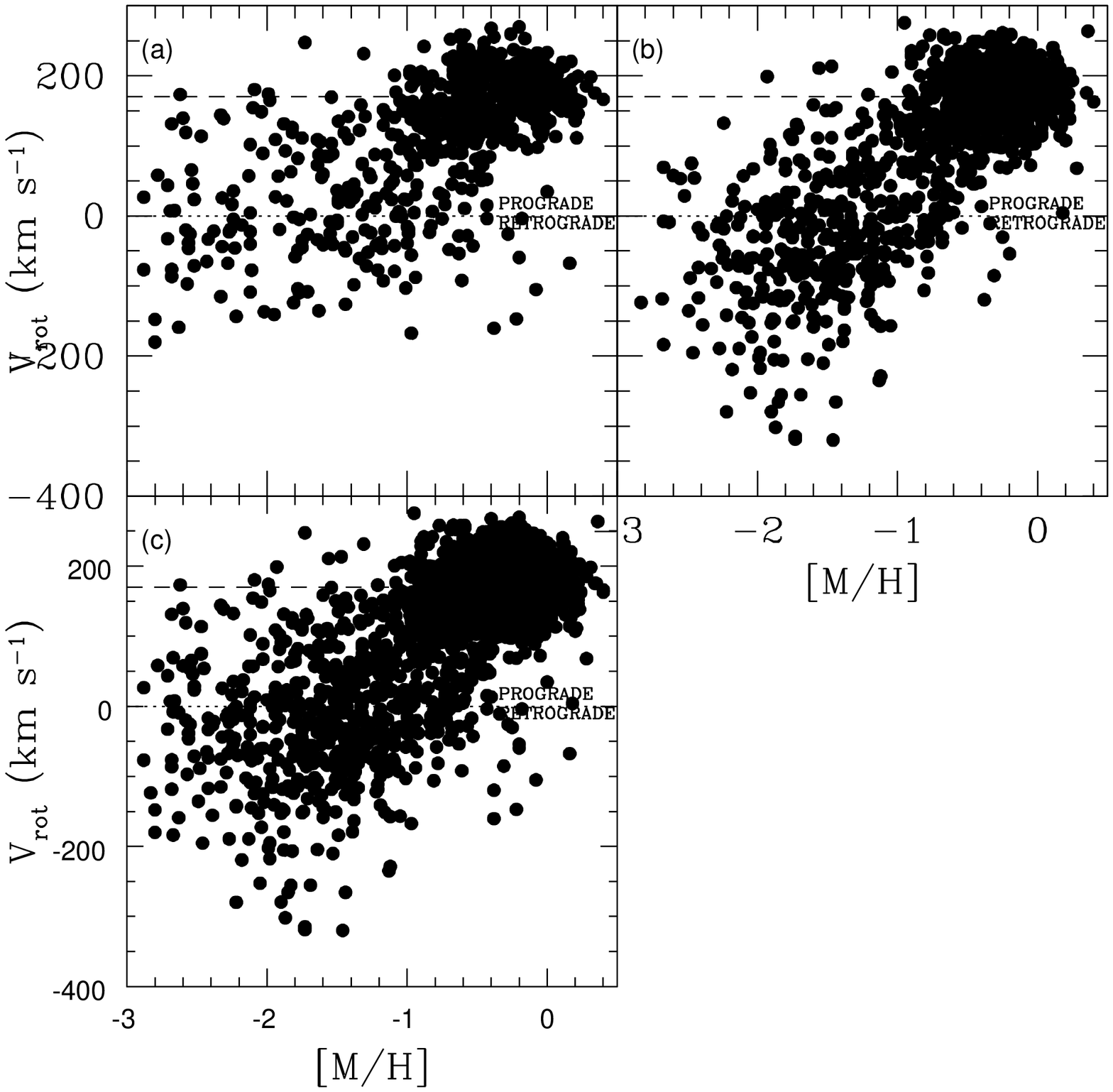}
%\epsffile{mwtd_arifprob.ps}
%\epsfxsize=18truecm
\caption{ Stars from Arifyanto et al.'s catalogue (panel a), Schuster's catalogue (panel b) and
the combined sample from the two catalogues (panel c) are shown in the ([M/H], $V_{\rm rot}$)
plane. The broken line at V$_{\rm rot}$ = 170 km s$^{-1}$ represents the mean velocity of
the [Fe/H] $\geq$ $-1$ thick disc stars, a drift velocity of 50 km s$^{-1}$ with respect to the
V$_{\rm rot}$ = 220 km s$^{-1}$ of the LSR. The dotted line at V$_{\rm rot}$ = 0 km s$^{-1}$ divides
stars in prograde from retrograde motion about the Galactic center.}

\end{figure}

\subsection{The Arifyanto et al. catalogue}

Arifyanto et al.'s (2005) sample 
of 742 subdwarfs is drawn from the proper motion selected samples
gathered and discussed by Carney et al. (1994, here CLLA).  
Arifyanto et al. adopted CLLA's radial velocities and spectroscopic metallicities 
but their
photometric parallaxes were  replaced by either the {\it Hipparcos}
measurement or a  photometric parallax for those
stars for which there was no {\it Hipparcos} accurate measurement.
Proper motions (and parallaxes) were taken from the
{\it Astrometric Catalog TYC2+HIP} (Wielen et al. 2001). The Tycho-2
proper motions represent about a factor of ten improvement in  accuracy
over the values adopted by CLLA.

Arifyanto et al.'s primary sample of 483 CLLA stars has kinematics from
TYC2 proper motions, {\it Hipparcos} parallaxes, and CLLA's
radial velocities. Their secondary sample of 259 stars uses
photometric parallaxes obtained from the 
comparison between CLLA's photometric parallaxes and
{\it Hipparcos} parallaxes for those stars with an accurate
{\it Hipparcos} parallax. The most metal-poor stars, say [Fe/H] $\leq  -1$, have
distances and space velocities determined predominantly from photometric
parallaxes, a result traceable to the decline in the stellar density distribution
with decreasing [Fe/H]. (This characteristic also applies to Schuster et al.'s
catalogue.) 

Distribution functions for the considered (primary plus secondary) sample
are shown in Figure~1 (shown by solid line). The sample is dominated by old thin and thick disc
stars (top left panel in Figure~1). Three panels show the distribution
functions for Galactic rotational velocity V$_{\rm rot}$ 
(= V$_{\rm LSR}$ + 220 km s$^{-1}$) in three metallicity bins. The entire sample
is shown in the ([M/H], V$_{\rm rot}$) plane in Figure~2a.

The CLLA sample was developed in two steps. Carney \& Latham (1987)
selected high proper motion stars from the Lowell and Luyten catalogues
by criteria designed to isolate F, G, and early K main sequence
stars. An additional criterion was imposed to select the `dynamically hottest'
stars, i.e., stars had to have at least two-thirds of their
(high) proper motion directed perpendicular to the disc or radially in the
Galactic frame. Obviously, application of this criterion reduces the
opportunity for thick disc stars to be chosen from the proper
motion catalogues. CLLA removed this kinematic bias
criterion and added about 500 (high proper motion) stars. 

Arifyanto et al. note that the
CLLA sample contains 1464 stars of which kinematics are given for 1269, 
metallicity for 1261, and radial velocity for 1447, and about 15\% belong
to binary or multiple stellar systems. Arifyanto et al.'s sample of
742 (483 primary and
259 secondary) stars  comprise about half of the CLLA
sample.

Our interest in Arifyanto et al.'s catalogue was piqued by their
conclusion
that in the metallicity range $-1.6 \leq$ [Fe/H] 
$\leq -1$ there are `a significant number of subdwarfs with disklike
kinematics', i.e., metal-weak thick disk stars. This conclusion and  procedures for 
identifying metal-weak thick disk stars are discussed below.

\subsection{The Schuster et al. catalogue}

Schuster et al. (2006) present a compilation of $uvby\beta$ photometry and
complete kinematic data for 1533 high-velocity and metal-poor stars.
When known and suspected binaries, probable variables, and stars with
lower quality photometry are excluded, the final database includes
1211 stars. Schuster et al.'s compilation draws on multiple sources of
high proper motion stars and metal-poor stars. Sources include the
CLLA catalogue, a similar one by Sandage \& Fouts (1987) which has a
35\% overlap with CLLA's catalogue, and other lists from the literature
and private submissions.

Schuster et al.'s catalogue
includes Str\"{o}mgren photometric estimates of metallicity [M/H], distances from
{\it Hipparcos} parallaxes when the error is less than 10\% 
or alternatively from 
calibrations based primarily on {\it Hipparcos} parallaxes, and
Galactic $U,V,W$ velocities from published proper motions and radial
velocities. Using the catalogue's astrometry, we have recomputed
(U$_{\rm LSR}$, V$_{\rm LSR}$, W$_{\rm LSR}$) for the sample stars (see above). 

Distribution functions for Schuster et al.'s sample are shown 
in Figure~1 (dotted line) in the
same format as for the Arifyanto et al.'s sample. Figure~2b shows clearly that the sample of halo stars
([M/H] $\leq$ $-$1.6) has a mean V$_{\rm rot}$ that is distinctly retrograde (V$_{\rm rot} < 0 $
km s${-1}$) whereas the sample in Figure~2a has a mean V$_{\rm rot}$ $\approx$ 0 km s$^{-1}$.
Sources other than the CLLA catalogue apparently provide significant
difference between the Arifyanto et al. and Schuster et al.
catalogue. 

Schuster et al. recognize three stellar populations based on 
their $V_{\rm rot}$ - [M/H] diagram: the old thin disc,
the thick disc, and the halo with   mean metallicities  and
dispersions  of ($<$[M/H]$>$,$\sigma_{[M/H]}$) = $(-0.16,0.14)$,
$(-0.55,0.18)$, and $(-1.40,0.60)$, respectively. Within the thick disc,
two components are suggested with different mean metallicities
([M/H] of $-0.4$ and $-0.7$),
rotation velocities, ages and velocity dispersions.
The question of metal-weak thick disk stars is not raised by
Schuster et al.

\subsection{ The combined catalogue}

By combining Arifyanto et al.'s sample of 742 stars with Schuster et al.'s
sample of 1211, we increase the sample to 1713 stars; there are 240 stars
in common to the two samples. 
Our recalculation of the space velocities shows that the two
samples give very similar results for the 240 stars in common.
The two catalogues have been merged without
adjustment of their space velocities as recalculated by us.
Distribution functions for the combined sample are shown
in the bottom four panels of Figure~1 with Figure~2c illustrating 
the sample in the ([M/H], V$_{\rm rot}$) plane.

Since the metallicities are obtained from different techniques, we
examine their consistency. CLLA obtain [M/H] from comparison of
(generally) low S/N high-resolution spectra of a narrow interval
around the Mg\,{\sc i} b lines (Carney et al. 1994). Schuster et al.
derive [Fe/H] from Str\"{o}mgren 
indices. The mean difference, in the sense of (Arifyanto $-$ Schuster)
is $-0.11\pm0.02$ dex. We have
corrected [M/H] values from Arifyanto et al. so that all the [M/H] values
in the combined sample are on the Schuster's scale.  

\subsection{Are there  local metal-weak thick disc stars?}

In principle, the thin disc, thick disc, and halo populations in the
solar neighbourhood may be defined kinematically by their
mean space velocities and  velocity dispersions.
At the metallicity of the metal-weak thick disk stars (i.e., [Fe/H] $\leq -1$),
contamination by the thin disc is quite negligible and
ignored here. Studies of local thick disc stars with [Fe/H] $\geq -1$ show that
the stars in the mean have a rotational velocity less than the
circular velocity of 220 km s$^{-1}$ by about 50 km s$^{-1}$ with
a dispersion $\sigma_{\rm V}$ of about 50 km s$^{-1}$. 
The mean velocities with respect to LSR
in the $U$ and $W$ directions are close to zero with dispersions
$\sigma_{\rm U}$ $\simeq$ 67 km s$^{-1}$ and  $\sigma_{\rm W}$
$\simeq$ 42 km s$^{-1}$. 
Defining the kinematic criteria for halo stars in the 
solar neighbourhood is complicated by the suggestion that the
halo consists of two components: 
an inner halo with
a metallicity centered on [Fe/H] $\approx$ $-$1.6 and a small net prograde velocity, and
an outer halo with a metallicity centered on [Fe/H] $\approx$ $-$2 and a net retrograde motion
(see, for example, Carollo et al. 2007).
The halo is  considered
to have no net motion in the $U$ and $W$ directions but the
mean velocity in the direction of Galactic rotation is here assumed
to correspond to $V_{\rm LSR}$ = $-$220 km s$^{-1}$.
Velocity dispersions for the halo are
about twice  
thick disc values.

In principle, a separation of  metal-weak thick disk stars from halo stars 
could be based on
relative probabilities following
the method  applied by Mishenina et al. (2004) and
Reddy et al. (2006) to isolate 
samples of local thick disc stars with [Fe/H] $> -1$ in the presence of
thin disc (and halo) stars.
In practice, application of the method
to metal-weak thick disk stars is severely hindered by three
issues: (i) the velocity characteristics of the metal-weak thick disk must be
extrapolated from those measured for the dominant thick disc ([Fe/H] $\geq$ $-$1)
population, (ii) the relative population of metal-weak thick disk and halo stars is
not known {\it a priori}, and (iii) selection effects 
permeate the chosen catalogues  with ill-understood effects on their
consequences for the metal-weak thick disk to halo mixture.

Evidence for metal-weak thick disk stars is suggested by Arifyanto et al. but occasions
no comment by Schuster et al. from their much larger data set. 
To address this
odd circumstance, we  discuss  the two catalogues separately,
and the combined one. This discussion brings to the fore the 
kinematical definition of a thick disc star and the possible dependence
of this definition on metallicity.

\subsubsection{Arifyanto et al.'s catalogue}

Arifyanto et al.'s  claim that `In the intermediate metallicity
range ($-1.6 \leq$ [Fe/H] $\leq -1$), we find a significant number of
subdwarfs with disklike kinematics' (i.e., metal-weak thick disk stars)
is made from the histograms reconstructed in Figure~1 (solid line).\footnote{Arifyanto et al.
 provide separate  histograms for 
samples with an accurate {\it Hipparcos} parallax and with a photometric
parallax. These samples provide similar-looking histograms and we chose not
to make a distinction by source of parallax in order to increase the
number of contributing stars.}   
 The top left histogram
shows the metallicity distribution for the sample. The other three
histograms show the distribution with $V_{\rm rot}$ for three metallicity
bins. The well sampled [Fe/H] $> -1.0$ histogram is peaked at about the expected
velocity of the thick disc and with a width anticipated from published
measures of the dispersion ($\sigma_V$). (Old thin disc stars may also
contribute to this histogram.) The few stars with [Fe/H] $< -1.6$ 
provide a histogram that is approximately symmetrical about
$V_{\rm rot} = 0$ with a width expected for stars with halo kinematics.
The intermediate metallicity bin, $-1.6 \leq$ [Fe/H] $\leq -1$
is neither symmetrical about $V_{\rm rot}$ = 0 of the halo or V$_{\rm rot}$ = 170 
of the thick disc but 
might be resolved into  halo  and thick disc contributions. It is this thick disc
contribution that yields the metal-weak thick disk stars recognized by Arifyanto et al. 
However, the halo contribution requires a velocity
dispersion less than that indicated by the bin with [Fe/H] $< -1.6$, and the
thick disc contribution calls for a distribution centred on a velocity
$V_{\rm rot} \simeq 110$ km s$^{-1}$, a velocity  lag substantially larger
than that associated with the thick disc at [Fe/H] $> -1$. 
Furthermore, isolation of these metal-weak thick disk candidates is based on very small
samples.
The asymmetrical 
histogram for $-1.6 \leq$ [Fe/H] $\leq -1$ 
was Arifyanto et al.'s evidence for subdwarfs with disclike
kinematics.

The full sample is shown in Figure~2a where we plot $V_{\rm rot}$ against [M/H].
A first impression is that the
stars may be separated into two groups. This is not a novel conclusion.
Carney et al. (1996) in describing their version of Figure 2 composed of
1022 stars wrote `it almost appears that there are (at least) two quite
distinct populations: one metal-poor and dynamically hot, and the other
metal-rich and dynamically cool (disk-like), and that  there is no
obvious sign of evolution of one into the other'. The disc-like
group with [Fe/H] $\geq -1$
 consists  of thick disc stars with  contamination likely by old thin
disc stars.
The metal-poor dynamically-hot stars are taken to
 belong to the halo.
Sandage \& Fouts (1997) in a similar diagram using ultraviolet excess $\delta$
in lieu of [Fe/H] remarked on the `abrupt break in the number of stars
at $\delta = 0.15$, with a much thinner distribution for $\delta > 0.15$';
$\delta > 0.15$ corresponds to [Fe/H] $< -0.8$ with the authors'
calibration. Sandage \& Fouts note that the distribution of W velocities
indicates that the densely and thinly populated distributions correspond
to the thick disc and the halo, respectively. 

Arifyanto et al.'s sample of metal-weak thick disk stars fall
in Figure~2a across the interval $-1.6 \leq$ [Fe/H] $\leq -1.0$. 
Much more striking, however, is the paucity of stars in what 
might be termed the expected belt for metal-weak thick disk, that is stars within
the strip centred on $V_{\rm rot}$ = 170 km s$^{-1}$ with a width
set by the dispersion $\sigma_V = 50$ km s$^{-1}$. For [Fe/H $> -0.5$,
the strip is densely populated. 
At [Fe/H] $<-1$, there are stars in the lower half of the
strip but almost none in the upper half. Therefore, if the kinematical
definition of the thick disc provided by local stars with
[Fe/H] $> -0.5$ applies to more metal-poor thick disc stars, 
and if selection effects have
not favoured stars in the lower half over those in the upper half of the strip,
this
catalogue contains almost no metal-weak thick disk stars.\footnote{Inspection 
of the thick disc
selection made by Reddy et al. (2006) show that their $V_{\rm rot}$
values are not symmetrically distributed about $V_{\rm rot} = 170$ km s$^{-1}$. 
The reason for this is very simple; it is only stars with $V_{\rm rot}$
$\leq 170$ for which the probability of belonging to the
thin disc drops to low values and  that of belonging to the thick disc
increases to a significant level.} 

\begin{figure} 
\includegraphics[width=9cm]{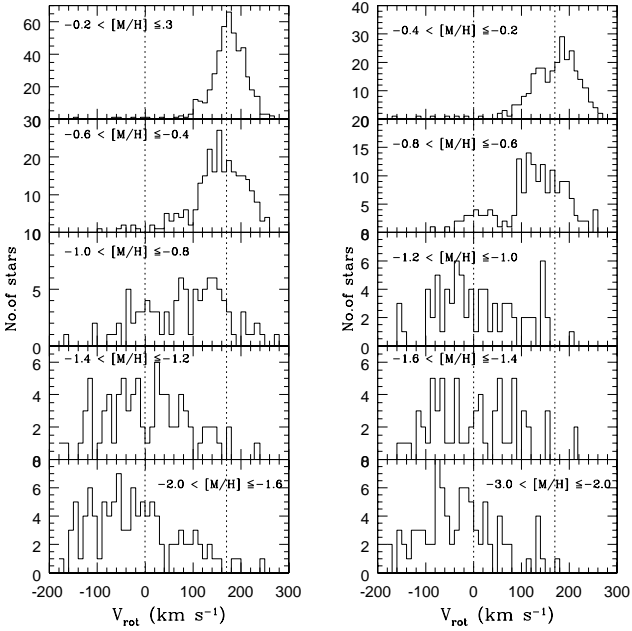}
%\epsffile{mwtd_schprob.ps}
%\epsfxsize=18truecm
\caption{ Histograms constructed from the combined catalogue for
metallicity bins showing the distribution of stars according to
V$_{\rm rot}$. Dotted lines are drawn for V$_{\rm rot}$ = 170 km s$^{-1}$, 
the mean velocity of thick disc stars with 
[Fe/H] $\geq$ $-$1.0, and V$_{\rm rot}$ = 0 km s$^{-1}$. }

\end{figure}

\subsubsection{Schuster et al.'s  catalogue}

Schuster et al.'s sample is shown in Figure~2b. 
There are differences between Figures~2a and 2b.
The mean $V_{rot}$ for the metal-poor stars in Figure~2b
is obviously  retrograde, as Schuster et al. note. 
The paucity of metal-weak stars in the strip centred on
$V_{\rm rot}$ = 170 km s$^{-1}$ is more obvious in Figure~2b than in Figure~2a. 
These differences highlight how (different) selection effects compound
the difficulty of identifying metal-weak thick disk candidates. 

Three groupings among stars with [Fe/H] greater than about $-1$ were
noted by Schuster et al. : (i) A concentration of old thin disc stars 
at ([Fe/H] in dex,$V_{\rm rot}$ in km s$^{-1}$) = ($-0.2$,176), (ii) a group of thick disc
stars at ($-0.4,151$), and (iii) a second group of thick disc stars at ($-0.7, 119$).\footnote{Schuster
et al.'s rotational velocities have been decreased by 9 km s$^{-1}$, as noted above.}
Schuster et al. suggest that the more metal-poor of the two thick disc groups is older by
about 3 Gyr.
The possible existence of two thick disc groups complicates the identification of metal-weak thick disk  stars.
Two obvious complications may be mentioned.
 Perhaps, the age-metallicity difference between the two groups implies an evolutionary
sequence for the thick disc  and then the  more metal-poor thick disc stars might
have a even lower $V_{\rm rot}$ which would place them among halo stars. Alternatively,
metal-weak thick disk stars might be associated with one or both of the groups of thick disc. Such
metal-weak thick disk stars, especially those from the more metal-poor group, would fall among halo stars
at the positive limits of the $V_{\rm rot}$ distribution for halo stars.

\begin{figure}
\includegraphics[width=9cm]{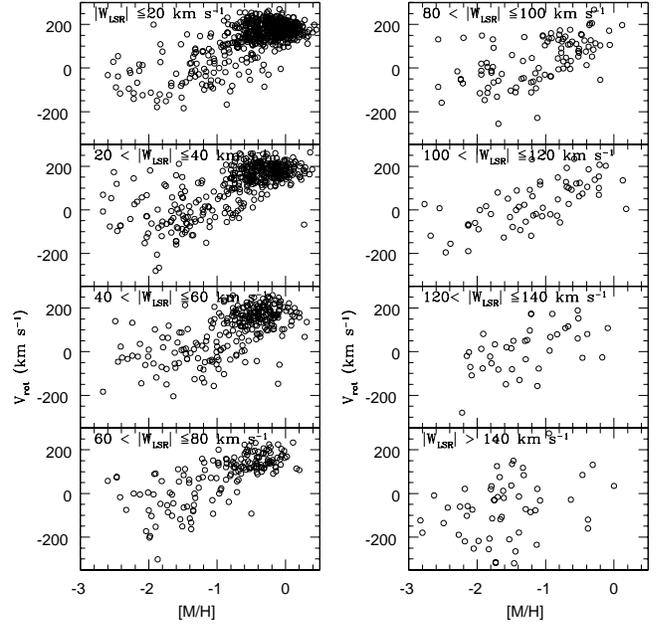}
%\epsffile{mwtd_schprob.ps}
%\epsfxsize=18truecm
\caption{ Stars from the combined catalogue are shown in the
 ([M/H],V$_{\rm rot}$) plane
in  bins of 20 km s$^{-1}$ in $|W_{\rm LSR}|$ velocity.}

\end{figure}

\begin{figure}
\includegraphics[width=9cm]{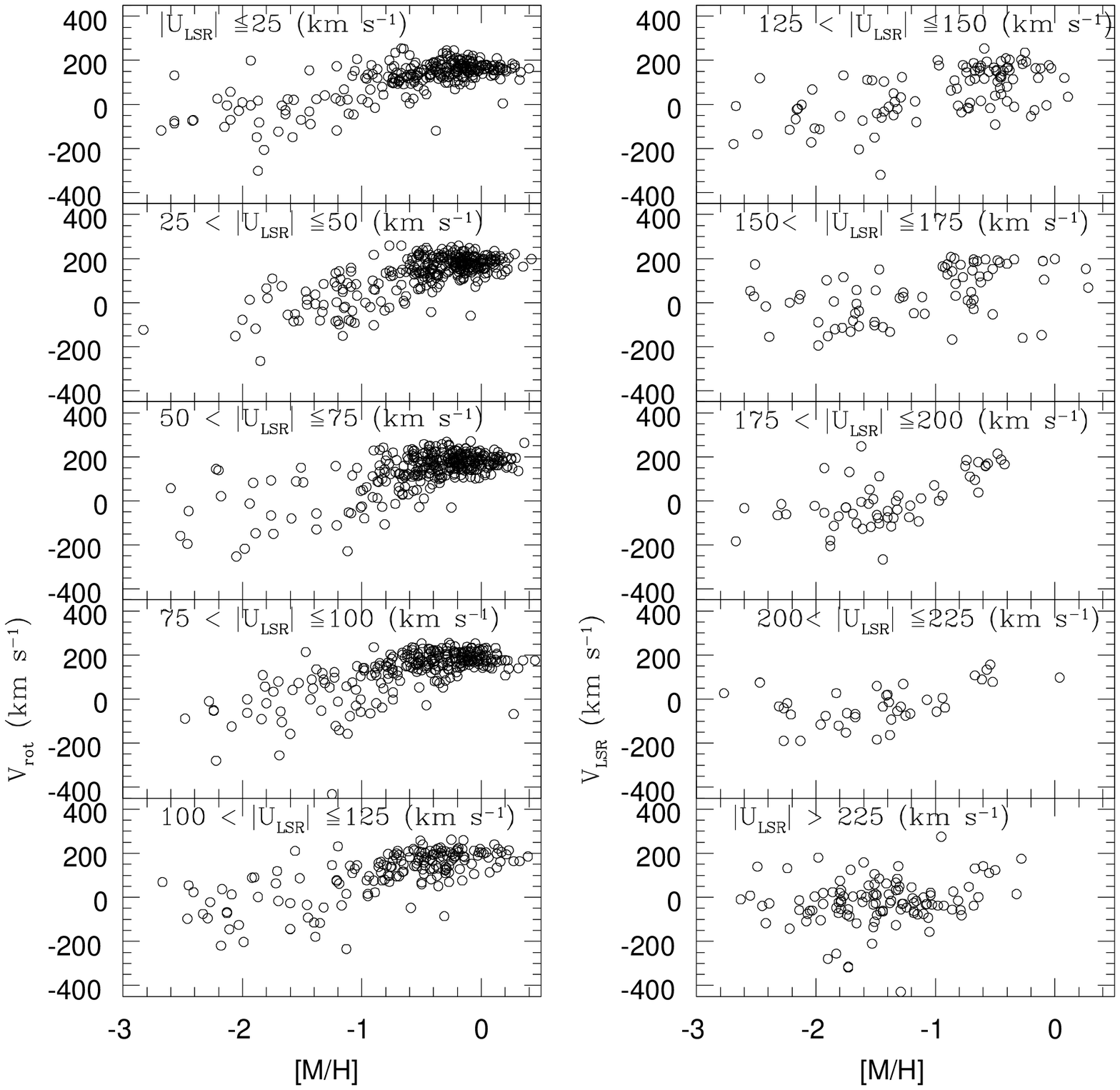}
%\epsffile{mwtd_schprob.ps}
%\epsfxsize=18truecm
\caption{Stars from the combined catalogue are shown
in the ([M/H],V$_{\rm LSR}$) plane
in  bins of 25 km s$^{-1}$ in $U_{\rm LSR}$ velocity. }

\end{figure}

\begin{figure}
\includegraphics[width=9cm]{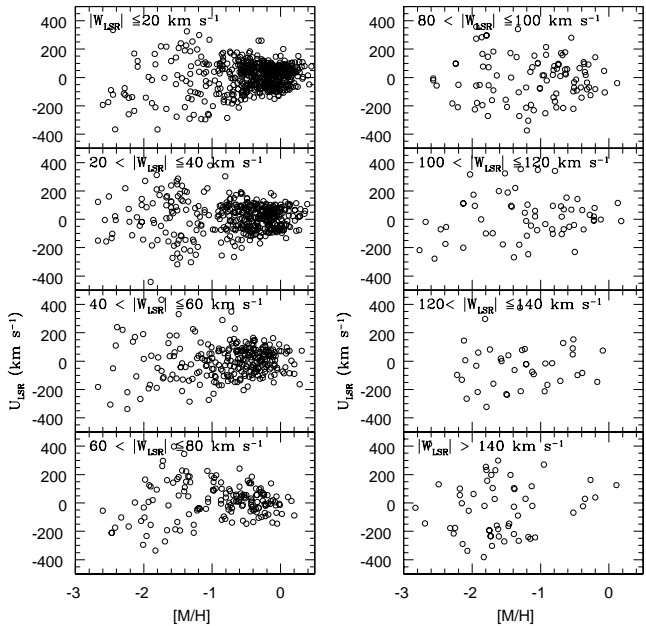}
%\epsffile{mwtd_schprob.ps}
%\epsfxsize=18truecm
\caption{The ([M/H], U$_{\rm LSR}$) plane as populated by the combined
catalogue for various $|W_{\rm LSR}|$ bins.}

\end{figure}

\subsubsection{The combined catalogue}

Although the ill-defined selection effects must not be forgotten,
the sample of 1713 stars promises to
reveal insights into the makeup of the metal-poor population less clearly glimpsed
from the Arifyanto et al. sample of 742 stars and Schuster et al.'s sample of
1211 stars. The ([M/H],$V_{\rm rot}$) diagram is shown in Figure~2c.
Interpretation of Figure~2c is assisted by constructing histograms
for bins in [M/H] (Figure~3). 
 Scanning the histograms from the most metal-rich
to the most metal-poor, we see the dominance of the thick disc at $V_{\rm rot}$
$\simeq$ 170 km s$^{-1}$ most distinctly and not surprisingly in the bin for
$-0.6 <$ [M/H] $< -0.4$ and the appearance of Schuster et al.'s second group
in the    $-0.8 <$ [M/H] $< -0.6$ and $-1.0 <$ [M/H] $< -0.8$ bins. The bin
$-0.4 <$ [M/H] $< -0.2$ bin, as might be expected, appears to be a mix of thin
and thick disc stars. 
Oddly, the histogram for $-0.2 <$ [M/H] $< +0.3$, the interval
providing the most metal-rich
stars, is peaked at the canonical velocity of the thick disc stars.
This is, perhaps, due to the selection effects emphasising 
high velocity stars made in the construction of Schuster's catalogue.
It may be interesting to note that thick disk
may extend beyond $-$0.3 towards higher metallicity.
The bins for [M/H] $< -1.0$ are filled with halo stars
with no apparent signature of metal-weak thick disk stars, i.e., the histograms are symmetrical
to within the noise about a velocity V$_{\rm rot} \leq 0 $ km s$^{-1}$. 
The mean $V_{\rm rot}$ declines
with decreasing [M/H] but this is entirely attributable to Schuster et al.'s
selection of stars. 

A dissection of Figure~2c by [M/H] is shown in Figure~4 where it is
reproduced for different intervals of vertical velocity $|W_{\rm LSR}|$.
 Here, one sees, as anticipated, the relative
mix of disc to halo stars decline as $|W_{\rm LSR}|$ increases. The
thick disc is seen as a distinct concentration for $|W_{\rm LSR}| < 100$
km s$^{-1}$, as expected because $\sigma_W \simeq 40$ km s$^{-1}$ for the
thick disc.  Figure~2c is resolved into $|U_{\rm LSR}|$ bins in Figure~5 which seems
to show, even more clearly than Figure~4, the sharp transition at [M/H] $\approx$ $-$1.
The disc component is traceable to 
$|U_{\rm LSR}|$ $\approx$ 200 km s$^{-1}$ for [Fe/H] $\approx$ $-$0.8. 
Velocities $|W_{\rm LSR}|$ $\leq$ 100 km$^{-1}$ and $|U_{\rm LSR}$ $\leq$ 140 km s$^{-1}$ 
ensures a partial exclusion of  halo stars
and the thin disk is excluded by [M/H]. Below [M/H] = $-$0.8 population
of thin disk is significantly reduced compared to thick disk and halo.
Figure~6 may suggest that stars below [M/H] = $-$1.0 and with 
$|W_{\rm LSR}|$ $\leq$ 100 km s$^{-1}$ and $|U_{\rm LSR}|$ $\leq$ 140 km s$^{-1}$
are probably thick disk stars not  thin disk stars.

As noted above, calculation of a probability of a metal-poor star belonging to either
the thick disc or the halo is fraught with uncertainty. 
Consideration of Figures~2c, 4, and 5 suggests that
the criteria $V_{\rm rot} > 100$ km s$^{-1}$, 
$|U_{\rm LSR}|$ $< 140$ km s$^{-1}$ and $|W_{\rm LSR}| < 100$
km s$^{-1}$ may enhance the ratio of metal-weak thick disk stars to halo stars selected from
the combined catalogue. The limit on
$V_{\rm rot}$ is based on the the distribution of stars with [M/H]  $> -0.6$ in
Figure~2c. As noted earlier, this selection criterion for metal-weak thick disk requires that
the  $V_{\rm rot}$  distribution for metal-weak thick disk stars differ  from that
for the more metal-rich thick disc stars.  The $|U_{\rm LSR}|$ criterion
follows in part directly from Figure~5 but also from the simple
condition that a disc-like star, i.e., one in a roughly circular Galactic
orbit, cannot have a $|U_{\rm LSR}|$ velocity that is a large fraction of its rotational
velocity. The $|W_{\rm LSR}|$ velocity condition follows from Figure~4.

Application of the velocity criteria to the combined catalogue yielded
a list of stars that was further culled by declination and magnitude. A few stars
satisfying all conditions were subsequently dropped from the abundance analysis
because their absorption lines were broad or binarity was suspected. When the
limit [Fe/H] $<$ $-$1.0 was applied following the abundance analysis, the 
fourteen candidate
metal-weak thick disk stars in Table~1 met all criteria. The eight stars labeled as thick disc stars 
have [Fe/H] $\geq$ $-$1.0
but otherwise met the criteria. Twenty stars 
selected initially by a different set of criteria fail to satisfy the above
velocity criteria, we label them as hybrid metal-weak thick disk/halo stars. 
This selection is likely a mix of thick disc stars  and halo stars.
A comparison sample of 18 halo stars were selected based on probability criteria
with $P_{\rm halo}$ $\geq$ 0.95 (see for details Reddy et al. 2006).
Probability criteria includes star's $V_{\rm rot}$
as well as $W$, $U$ velocities and their dispersions.
The parameters of halo sample are
listed in Table~2. Stars selected for abundance analysis are highlighted in Figure~7.

\begin{figure*}
\includegraphics[width=12cm]{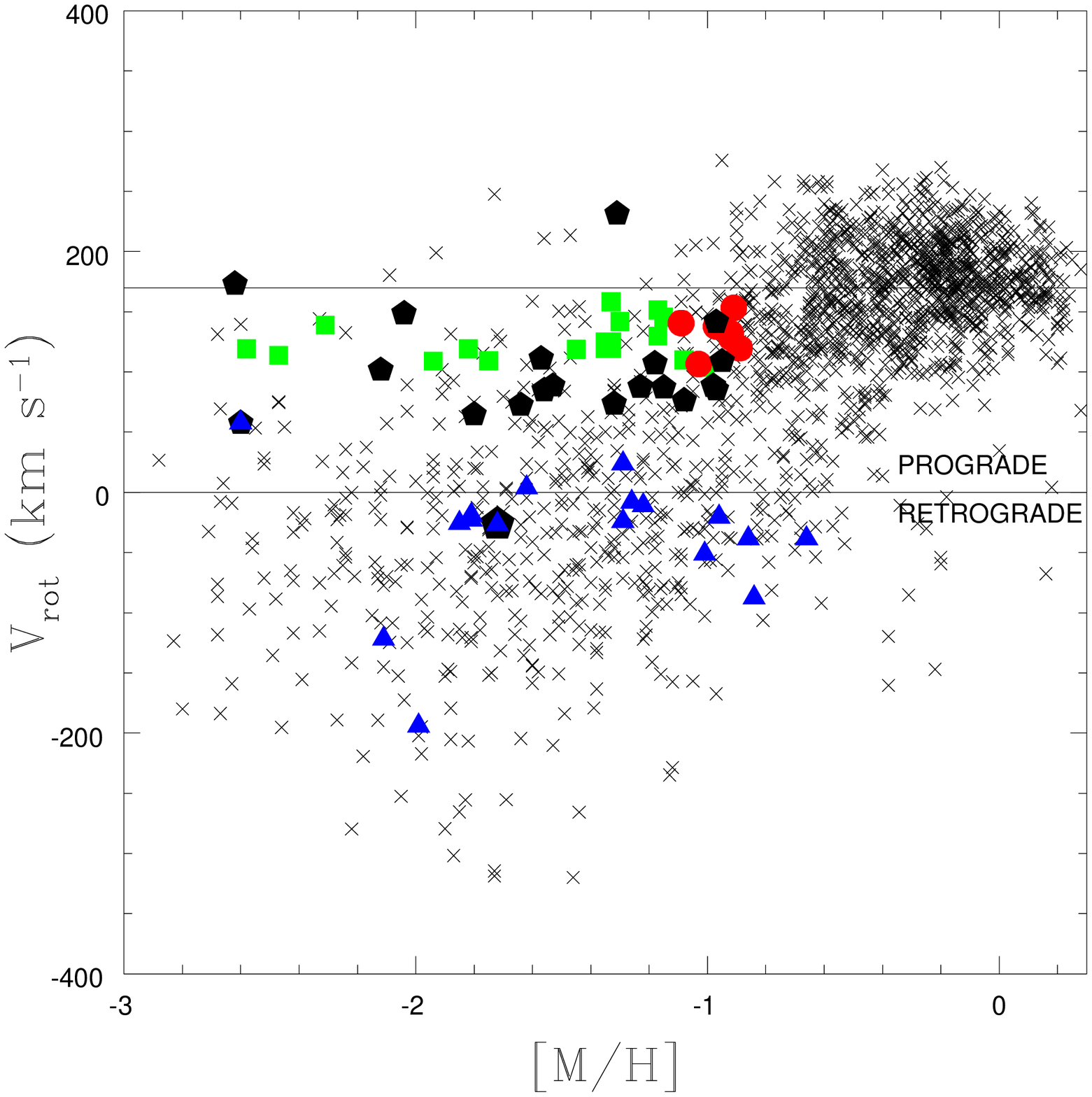}
%\epsffile{mwtd_schprob.ps}
%\epsfxsize=18truecm
\caption{ The ([M/H],V$_{\rm rot}$) plane for the stars from 
the combined catalogue (black crosses) showing the 14 metal-weak thick disk
candidates (green squares), 8 thick disc stars (red circles), 
20 hybrid stars (Thick disc/halo) (black pentagons)  from Table~1,
and the 18 halo comparison sample from Table~2 (blue triangles). Two stars
that have large radial velocity discrepancy are shown as two large symbols
(green square and black pentagon).}

\end{figure*}

\section {Observations}

Spectra were obtained 
using the 2dcoud\'{e} echelle spectrograph (Tull et al. 1995) fed by the
Harlan J. Smith 2.7-m telescope
of the W.J. McDonald Observatory.
 Spectra were recorded on a Tektronix 2048 $\times$ 2048 
CCD detector.
Spectral coverage is similar to that provided by observations contributing to
our earlier papers (Reddy et al. 2003, 2006).  
Present and earlier spectra were obtained with
a resolving power of $R = \Delta \lambda / \lambda \simeq$ 60,000. 
Standard reduction procedures within $IRAF$\footnote{IRAF is distributed by the National
   Optical Astronomy Observatories, which are operated by the
   Association of Universities for Research in Astronomy, Inc., under
   cooperative agreement with the National Science Foundation."}
   were used 
to convert the raw 2D-Spectra into 1D-spectra of relative flux versus wavelength and
to measure equivalent widths (W$_{\lambda}$) of the selected lines.

Radial velocity measurements were made for all stars. Our measurements
are in general in excellent agreement with values reported in the Arifyanto et al.
and Schuster et al. catalogues. In two cases, a substantial  difference
was noted: e.g., G141-47, has a radial velocity of $-32.7$ km s$^{-1}$
but Schuster et al.'s catalogue lists $-23.4$ km s$^{-1}$, and the halo star
HIP4754 has a radial velocity of $-69.8$ km s$^{-1}$  but Arifyanto et al.'s
catalogue lists $-86.9$ km s$^{-1}$. We suppose that these may be examples of
spectroscopic binaries. For all stars, our measured radial velocities were used 
with the parallax and proper motions from the catalogues to recompute the
LSR velocities.

\begin{table*}
\begin{center}
\caption{Atmospheric parameters, kinematic and orbital parameters 
for the 14 candidate metal-weak thick disk stars, 8 thick disc stars and 20 hybrid (thick/halo) stars.
Columns 1-9 are self explanatory. The entry (R$_{\rm v}$)
in column 6 is the heliocentric radial velocity and the values given in the bracket are
from the catalogues. 
The orbital parameters (R$_{\rm m}$, $e$, and Z$_{\rm max}$) are discussed in the text. The age
in Gyrs is given in column 13. 
The adopted $T_{\rm eff}{p}$ and log $g$ values based on photometry and
parallaxes  are given in column 3 and 5, respectively. Values that are adopted
from spectroscopic analysis 
are denoted by a superscript 's'.}  \end {center}
\begin{tabular}{rccclrrrrrrrrc}
\hline \hline
Star   &  [Fe/H] & $T_{\rm eff}{p}$ & $T_{\rm eff}{s}$ & log $g$ & $R_{\rm v}$  & $U_{\rm LSR}$ & $V_{\rm LSR}$ & $W_{\rm LSR}$ & $R_{\rm m}$ & $e$ &
$Z_{\rm max}$ & $\tau_{9}$  \\
          &          & (K)  & (K) & (cm s$^{-2}$) & \multicolumn{4}{c}{(km s$^{-1}$)}& (kpc)  & & (kpc)  & (Gyrs) \\
\hline
\multicolumn{12}{c}{Metal Weak Thick Disk Stars} \\
\hline
G242-071  &  $-$1.15 & 5978 & 6030 & 3.61 & $-$1  &  -118 & -89  &  47  & 6.75 &  0.529 &  0.479 &   13.1$^{+2.4}_{-4.8}$ \\
G172-058  &  $-$1.53 & 6131 & 6250 &  4.78 & $-$25 &  -93  & -110 &  19  & 6.03 &  0.572 &  0.163 &   ...                  \\
G071-055  &  $-$1.54 & 5717 & 5750 &  4.78 & $-$30 &  112  & -100 & -44  & 6.46 &  0.558 &  0.417 &  ...                   \\
G102-020  &  $-$1.17 & 5246 & 5340 & 4.63 & 20     & -4    &  -68  &   70  &  6.52 &   0.306 &   0.731 &   ...          \\
G146-076  &  $-$1.78 & 5063  &4960 & 4.11 & $-$113 &  51   &  -81  &   -90 &  6.43 &   0.389 &   1.166 &  ...                   \\
G062-052  &  $-$1.11 & 5271$^{s}$  &5450 & 4.50$^{s}$ & $-$47  &  50   &  -61  &   -83 &  6.86 &   0.315 &   0.892 &  20.5$^{+3.2}_{-4.6}$  \\
G166-045  &  $-$2.32 & 6016  & ... & 4.21 & 29     &  131  &  -102 &   23  &  6.71 &   0.587 &   0.242 & ...                    \\
G141-047  &  $-$1.33 & 6083  & 6080 &4.52 & $-$33  &  69   &  -104  &   -22 &  5.97 &   0.520 &   0.176 &   9.0$^{+3.0}_{-1.4}$ \\
          &          &       &      &     &($-$23)   &       &       &       &       &         &         &         \\
HD201891  &  $-$1.05  &  5853  & 5850& 4.09 & $-$43  &  102 &  -111 &   -51 &  6.16 &   0.577 &   0.516 &  18.1$^{+9.0}_{-13.8}$ \\
G188-022  &  $-$1.29  &  5950  & 6020& 4.26 & $-$93  &  139  &  -99 &   62  &  6.85 &   0.598 &   0.732 &    ...                 \\
G060-026  &  $-$1.14  &  5287  & 5300& 4.00$^{s}$ &  114   &  80   &  -116 &   89  &  5.90 &   0.572 &   1.124 &  10.9$^{+12.0}_{-7.9}$ \\
G014-033  &  $-$1.04  &  5285  & 5344& 4.81 & $-$91  & -99  &  -70  &   -90 &  7.02 &   0.445 &   1.253 &   ...                  \\
HD16031   &  $-$1.62  &  6076  & 6238& 4.01 & 25     &  37   &  -95  &   -30 &  5.72 &   0.528 &   0.272 &   16.5$^{+2.1}_{-8.4}$ \\
G090-003  &  $-$1.89  &  5727  & ... & 4.89 & 30     & -15  &  -106 &   -7 &  5.70 &   0.493 &   0.050 &   ...                  \\
\hline
\multicolumn{12}{c}{ Thick Disk Stars} \\
\hline
G084-037  &  $-$0.94 & 5953 & 5950&  4.23 & $-$13 &  18   & -89  &  95  & 6.18 &  0.394 &  1.249 &   13.8$^{+1.7}_{-2.3}$ \\
G113-022  &  $-$1.00 & 5601 & 5570&  4.25$^{s}$ & 56     &  34   &  -80  &   64  &  6.36 &   0.367 &   0.649 &   10.0$^{+9.3}_{-3.0}$ \\
G010-012  &  $-$0.64 & 4930  & 5035& 3.50$^{s}$ & 133    & -53   &  -91  &   96  &  6.24 &   0.434 &   1.274 &   ...                  \\
G236-082  &  $-$0.57 & 5622  & 5620& 4.81 & $-$71  & -81   &  -67  &   -96 &  7.01 &   0.378 &   1.320 &  7.5$^{+3.0}_{-1.4}$  \\ 
G066-051  &  $-$0.91 & 5234  & 5260 & 4.80 & $-$118 & -86   &  -79  &   -68 &  6.67 &   0.437 &   0.770 &  ...                   \\
G126-036  &  $-$0.78  &  5632  & 5620 & 4.78 & $-$86  &  81   &  -113 &   8   &  5.88 &   0.570 &   0.064 &  12.0$^{+4.8}_{-0.7}$  \\
G029-025  &  $-$0.84  &  5407  & 5300 & 4.63 & $-$87  & -102  &  -103 &   25  &  6.25 &   0.531 &   0.243 &   ...                  \\
G262-032  &  $-$0.87  &  5000  & 5100 &4.75$^{s}$ & $-$87  &  118  &  -82 &   74 &  6.99 &   0.502 &   0.920 &   ...                  \\
\hline
\multicolumn{12}{c}{Thick disc/Halo} \\
\hline
G172-016  &  $-$0.98 & 5621 &  5620 & 4.50$^{s}$ & $-$84 &  -88  & -146 &  32  & 5.37 &  0.722 &  0.278 &   ... \\
G058-025  &  $-$1.34 & 5936  & 6020 & 4.09 & 66     & -25   &  -135 &   19  &  5.23 &   0.636 &   0.141 &  18.6$^{+4.5}_{-1.4}$  \\
HD97916  &   $-$0.82 & 6238  & 6270 & 3.55 & 63     &  116  &  11   &   103 &  10.24 &  0.363 &   1.811 & 7.3$^{+1.7}_{-0.8}$    \\
G016-013  &  $-$0.89 & 5550  & 5680 & 4.00$^{s}$ & $-$51  & -27   &  -133 &   15  &  5.27 &   0.626 &   0.107 &  3.8$^{+9.2}_{-2.7}$   \\
G017-016  &  $-$0.77 & 5238  & 5357 & 4.77 & $-$162 & -140  &  -111 &   -19 &  6.63 &   0.634 &   0.168 &  ...                   \\
G139-049  &  $-$0.94 & 5344  & 5465 &4.87 & $-$94  & -26   &  -131 &   -7  &  5.28 &   0.623 &   0.056 &    ...                 \\
G204-049  &  $-$0.91 & 5151  & 5260 & 4.80 & $-$41  &  171  &  -77  &   -59 &  8.01 &   0.592 &   0.754 &   ...                  \\
G241-007  &  $-$0.75  &  5593  & 5648 & 4.80 & $-$112 & -46   &  -133 &   -21 &  5.31 &   0.640 &   0.165 &  11.5$^{+4.3}_{-2.5}$  \\
G143-017  &  $-$1.41  &  5435  & 5360 &4.82 & $-$191 & -143  &  -109 &   70 &  6.77&   0.627 &   0.867 &   ...                  \\
G019-025  &  $-$1.78  &  5009  & 5000 & 5.00$^{s}$ & $-$32  &  39   &  -129 &   -37 &  5.01 &   0.738 &   0.347 &   ...                  \\
G059-018  &  $-$1.13  &  5035  & 5050 & 3.60 & 33     & -92   &  -146 &   9   &  5.46 &   0.709 &   0.037 &   ...                  \\
G027-008  &  $-$1.31  &  5845  & 5845 & 4.74 & $-$53  & -87   &  -129 &   -97 &  5.74 &   0.647 &   1.419 &   12.4$^{+7.1}_{-1.6}$ \\
G271-162  &  $-$2.33  &  6009  & ... & 4.14 & 37     &  -162 &  -47  &   29  &  8.65 &   0.511 &   0.315 &   10.2$^{+0.0}_{-0.0}$ \\
G005-001  &  $-$0.91  &  5704  & 5620 &  4.52 & $-$22  &  44   &  -121 &   -82 &  5.73 &   0.533 &   0.877 &   18.5$^{+3.6}_{-2.4}$ \\
G123-009  &  $-$1.16  &  5379  & 5380 &4.74 & $-$22  & -104  &  -146 &   -19 &  5.53&   0.729 &   0.161 &   ...                  \\
G016-020  &  $-$1.59  &  5386  & 5550 & 4.94 &   171  &  180  &  -71  &   93  &  8.55 &   0.595 &   1.519 &   ...                  \\
G204-030  &  $-$0.83  &  5570  & 5570 &4.72 & $-$69  &  58   &  -131 &   47  &  5.42 &   0.631 &   0.434 &   ...                  \\
G037-026  &  $-$1.89  &  5918  & ... &4.41 & $-$140 &  167 &  -118 &   -60 &  7.00 &   0.691 &   0.721 &   ...                  \\
HIP96115  &  $-$2.66  &  5500  & ... &3.39 & $-$128 &  55   &  -161 &   -67 &  4.95 &   0.772 &   0.672 &   ...                  \\
HIP4754   &  $-$1.72  &  5518  & 5550 & 3.95 & $-$70  & -116  &  -246 &    5  &  5.08 &   0.921 &   0.041 &   ...                  \\
          &          &       &      &     &($-$87)   &       &       &       &       &         &         &         \\

\hline
\end{tabular}
\end{table*}

\begin{table*}
\begin{center}
\caption{Atmospheric parameters, kinematic and orbital parameters 
for the 18 halo stars.
Columns 1-9 are self explanatory. The entry (R$_{\rm v}$)
in column 6 is the heliocentric radial velocity.
The orbital parameters (R$_{\rm m}$, $e$, and Z$_{\rm max}$ are discussed in the text. The age
in Gyrs is given in column 13. 
The adopted $T_{\rm eff}{p}$ and log $g$ values based on photometry and
parallaxes  are given in column 3 and 5, respectively. The stars for which
spectroscopic parameters are used in the analysis are denoted by a superscript 's'.} \end {center}
\begin{tabular}{rcclrrrrrrrrc}
\hline \hline
Star   &  [Fe/H] & $T_{\rm eff}{p}$ & $T_{\rm eff}{s}$ &log $g$ & $R_{\rm v}$  & $U_{\rm LSR}$ & $V_{\rm LSR}$ & $W_{\rm LSR}$ & $R_{\rm m}$ & $e$ &
$Z_{\rm max}$ & $\tau_{9}$ \\
         &        &    (K)    &(cm s$^{-2}$)        & \multicolumn{4}{c}{(km s$^{-1}$)}& (kpc)  & & (kpc)  & (Gyrs) \\
\hline
HIP3026  &  $-$1.22  &  5982  & 6050 & 4.77 & $-$47  &  145  &  -231 &  -33  & 91.05 &   0.997 &   9.069 &   15.2$^{+2.8}_{-3.5}$ \\
HIP3430  &  $-$1.81  &  6129  & ... &3.75 & $-$121  & -137  &  -243 &   51  &  5.32 &   0.936 &   0.534 &   16.3$^{+3.0}_{-3.5}$ \\
HIP10449 &  $-$0.89  &  5648  & 5620 & 4.40 &    28  & -196  &  -196 &   65  &  6.88 &   0.952 &   1.043 &   ...                   \\
HIP15904 &  $-$1.38  &  5746  & 5746 & 4.50$^{s}$ &    87  & -108  &  -262 &  -36  &  5.16 &   0.859 &   0.295 &   ...                  \\ 
HIP16404 &  $-$1.98  &  5241  & 5200 & 4.76 & $-$161  & -165  &  -414 &  -65  &  9.58 &   0.498 &   0.924 &   ...                 \\    
HIP38541 &  $-$1.62  &  5394  & 5394 & 4.74 & $-$234  & 270  &  -216 &  -81  &5111.13 &   0.999 &9850.129 &    ...                 \\   
HIP42887 &  $-$1.29  &  6793  & 6793 &3.51 &    58  &  -40  &  -244 &  -82  &  4.52 &   0.916 &   0.594 &   6.7$^{+0.1}_{-1.4}$  \\     
HIP80837 &  $-$0.63  &  5931  & 5900 & 3.79 & $-$47.4  &  96  &  -258 &  -70  &  4.88 &   0.866 &   0.475 &   10.9$^{+1.5}_{-1.5}$  \\    
HIP86321 &  $-$0.86  &  6017  & 6017 & 4.06 & $-$239.9  &  -74  &  -258 &  -44  &  4.83 &   0.865 &   0.343 &  11.7$^{+1.3}_{-1.0}$ \\     
HIP86443 &  $-$2.21  &  5977  & ...  &3.96 & $-$397.2  & -356  &  -246 &   86  & 21.15 &   0.984 &   3.571 &   ...                 \\    
HIP94449 &  $-$1.02  &  6194$^{s}$  & 5800 & 4.25$^{s}$ & $-$64.8  & 162  &  -307 &   59  &  6.62 &   0.731 &   0.652 &   6.3$^{+2.1}_{-1.8}$   \\   
HIP100568&  $-$0.96  &  5735  & 5622 & 4.70 & $-$171.3  & -146  &  -240 &  -66  & 13.56 &   0.981 &   1.343 & 14.2$^{+4.0}_{-4.8}$  \\    
HIP100682&  $-$2.11  &  5500  & ... &4.50$^{s}$ & $-$322.9  & -41  &  -342 &  -87  &  6.01 &   0.458 &   1.061 &    ...                 \\     
HIP100792 & $-$1.01   &  5997  & 5997 & 4.14 & $-$246.8  &  -66  &  -271 &  -23  &  4.92 &   0.808 &   0.171 &  13.1$^{+2.9}_{-1.7}$\\     
HIP105488 & $-$1.26   &  5928  & 5800 & 4.86 & $-$272.8  & -161  &  -228 &   62  &1605.62 &   0.999 &2847.656 &  ...                 \\     
HIP106924 & $-$1.81   &  5400  & 5370 &4.86 & $-$244.4  &  99  &  -239 &  115  & 157.70 &   0.998 &  49.767 &   ...                 \\    
HIP111195 & $-$1.63   &  5880  & 5950 &4.65 & $-$213.1  &  17  &  -265 &   61  &  4.65 &   0.824 &   0.578 &   16.9$^{+3.1}_{-0.9}$\\    
HIP117041 & $-$0.93   &  5357$^{s}$  & 5150 &4.00$^{s}$ &  $-$86.2  & -251  &  -245 &  -34  &  8.81 &   0.962 &   0.413 &   ...                \\
\hline
\end{tabular}
\end{table*}

\section{Abundance Analysis}

Abundance analysis follows procedures used by
Reddy et al. (2003,  2006). 
Abundances are derived using equivalent widths and
LTE model atmosphere grids
combined with the spectral analysis code MOOG (Sneden 1973).
Stellar atmospheric models (Kurucz 1998) 
computed for
LTE  with the convective overshooting option are employed in the analysis.
The  model for a particular $T_{\rm eff}$, log $g$, and metallicity
is extracted from the grid by linear interpolation.
The choice of  convective overshooting models over  convective models without
overshooting is discussed 
in Reddy et al. (2003) with justification for solar type stars
by (Castelli, Gratton, $\&$ Kurucz 1997). 
Stars
in our earlier surveys of thin disc and thick disc stars  and
the present metal-weak thick disk candidates are all main sequence stars within about $\pm$600 K of the solar T$_{\rm eff}$. 
There is of course a difference in metallicity between our stars and the Sun.
Differences in abundance ratios [X/Fe],
between convective and over convective models were earlier shown to be very small.

\begin{table*}
\caption{Iron abundance and abundance ratios [X/Fe] for 14 elements for the 14 candidate metal-weak thick disk stars, 8 thick disc star
and 20 hybrid stars (thick disc/halo).} 
\begin{tabular}{rrcrcrrrrrrrrrrr}
\hline \hline
Star&   [Fe\,{\sc i}/H]&    Na&    Mg&    Al&     Si&     Ca&    Sc&    Ti&    V &  Cr   &  Mn   &  Ni   &  Cu   &   Ba    &  Eu \\
\hline
\multicolumn{16}{c}{Metalweak Thick Disk} \\
\hline
G242-71&   -1.15&  -0.21&  0.22&  0.22&   0.18&   0.39& -0.02&  0.30&  0.17&  -0.13&  -0.63&   0.00& -0.57&   0.50&  0.79\\
G172-58&   -1.53&   ...&  0.29&   ...&   0.46&   0.35&  0.32&  0.38&  0.01&  -0.32&  -0.48&  -0.22& -0.37&  -0.02&  0.31\\
G71-55&    -1.54&   0.00&  0.08&  0.40&   0.16&   0.24&  0.34&  0.220& 0.07&  -0.37&  -0.53&  -0.17& -0.53&   0.23&  0.54\\
G102-20&   -1.17&  0.02&   0.27&   0.33&   0.24&   0.22&  0.26&   0.19&  0.04& -0.16&  -0.45& -0.09&  -0.10&    -0.08&   0.27\\
G146-76&  -1.78&   ...&   0.19&  ...&   0.28&   0.17&   0.54&   0.10& -0.06& -0.27&  -0.76& -0.12& -0.96& 0.04&   0.43\\
G62-52&  -1.11&  0.12&   0.28&   0.43&   0.35&   0.25&   0.29&   0.19& -0.07&  -0.12& -0.49& -0.03& -0.23& -0.03&  0.49\\
G166-45&  -2.32&   ...&   0.22&    ...&    ...&    ...&   0.38&    ...&  0.18&   ...&   ...&   ...&  ...& -0.51&   1.02\\
G141-47&  -1.33&  0.06&   0.34&   0.45&   0.33&   0.36&   0.15&   0.26&  0.56&  -0.27&  -0.56& -0.08&   ...&   0.16&   0.59\\
HD201891&  -1.05&  0.14&   0.35&   0.24&   0.16&   0.20&  0.05&   0.22&  0.09& -0.14&  -0.43&  0.04&  -0.25&   -0.05&  0.07\\
G188-22&  -1.29&  0.02&   0.31&   0.26&   0.25&   0.31&  0.14&   0.32&  0.14& -0.17&  -0.40& -0.06&  -0.60&     0.36&  0.29\\
G60-26&  -1.14& -0.32&   0.21&   0.09&   0.15&   0.23&  0.06&   0.15& -0.15& -0.14&  -0.44& -0.13& -0.76&    0.18&  0.48\\
G14-33&   -1.04& -0.05&   0.22&   0.33&   0.21&   0.23&  0.21&   0.23&  0.11& -0.21& -0.39&  0.05& -0.09&    0.02&  0.27\\
HD16031&   -1.62& -0.04&   0.32&   ...&   0.35&   0.32&   0.10&   0.36&  0.07&  -0.26&   ...& -0.09&   ...&-0.07&   0.21\\
G90-3&   -1.89&  0.14&   0.05&   ...&    ...&   0.24&   0.50&    ...& -0.14&  -0.49&   ...& -0.27&   ...&  0.00&   0.67\\
\hline
\multicolumn{16}{c}{Thick Disc Stars} \\
\hline
G84-37  &  -0.94&  0.06&   0.30&    ...&   0.37&   0.29&  0.09&   0.33&  0.11& -0.02&  -0.35& -0.04&  -0.52&   0.20&   0.28\\
G113-22&   -1.00& -0.05&   0.29&   0.27&   0.27&   0.24&  0.29&   0.24& -0.02& -0.11&  -0.66&  0.03&  -0.33&      0.49&   0.35\\
G10-12&  -0.64&  0.19&   0.43& 0.48&   0.31&   0.29&   0.13&   0.28&  0.16&  0.03&  -0.32&  0.24&  0.06&    ...&...\\
G236-82&  -0.57& -0.01&   0.23& 0.26&   0.17&   0.14&   0.30&   0.17&  0.07& -0.09&  -0.27&  0.06& -0.07&   -0.02&   0.17\\
G66-51&  -0.91& -0.03&   0.20&   0.22&   0.21&   0.22&   0.26&   0.25&  0.09&  -0.07& -0.39&  0.03&  0.01&    -0.05&  0.37\\
G126-36&  -0.78&  0.02&   0.21&   0.21&   0.16&   0.20&  0.12&   0.14&  0.10& -0.07&  -0.41& -0.02&  -0.10&     0.71&  0.31\\
G29-25&  -0.84&  0.06&   0.31&   0.29&   0.23&   0.33&  0.16&   0.35&  0.08& -0.02&  -0.35&  0.06& -0.03&    0.01&  0.24\\
G262-32&  -0.87&  0.05&   0.27&   0.39&   0.31&   0.33&  0.19&   0.39&  0.07&  0.02&  -0.26&  0.08&  0.10&    0.70&  0.31\\
\hline
\multicolumn{16}{c}{Thick Disc/Halo} \\
\hline
G172-16&   -0.98&  -0.11&  0.03&  0.07&   0.18&   0.13&  0.09&  0.09& -0.02&  -0.37&  -0.70&  -0.14& -0.94&   0.31&  0.50 \\
G58-25&   -1.34&  0.06&   0.19&   0.33&   0.22&   0.39&  0.02&   0.28&  0.09& -0.19&  -0.61& -0.08&  -0.50&    0.25&   0.14\\
HD97916&  -0.82& -0.07&   0.22& 0.08&   0.20&   0.32&   0.12&   0.15&  0.00& -0.24&  -0.56&  0.02& -0.64&  0.25&...\\
G16-13&  -0.89&  0.10&   0.32&   0.34&   0.31&   0.26&   0.23&  0.16& -0.07&  -0.19& -0.51& -0.02& -0.27&   0.29&  0.31\\
G17-16&  -0.77&  0.05&   0.27&   0.36&   0.31&   0.23&   0.12&   0.27&  0.08&  -0.11&  -0.36&  0.05&  0.04&   -0.06&   0.35\\
G139-49&  -0.94&  0.04&   0.21&   0.33&   0.30&   0.21&   0.25&   0.15&  0.09&  -0.11&  -0.41&  0.13&  0.01&  -0.07&   0.30\\
G204-49&  -0.91&  0.06&   0.25&   0.29&   0.28&   0.19&   0.32&   0.23&  0.06&  -0.10&  -0.39&  0.11& -0.14&   0.00&   0.38\\
G241-7&  -0.75&  0.03&   0.25&   0.38&   0.17&   0.28&  0.26&   0.32&  0.06& -0.12&  -0.42& -0.01&  -0.14&     0.15&  0.28\\
G143-17&  -1.41&   ...&   0.12&    ...&    ...&   0.21& -0.04&   0.17&  0.00& -0.09&  -0.54& -0.24& -0.51&   -0.01&  0.39\\
G19-25&   -1.78&   ...&   0.26&    ...&   0.28&   0.30&  0.02&   0.30&  0.15& -0.08& -0.76& -0.08&  ...&   -0.17&  0.71\\
G59-18&   -1.13& -0.17&   0.29&    ...&   0.23&   0.23&  0.14&   0.11& -0.16& -0.15& -0.59& -0.08& -0.50&   0.21&  0.34\\
G27-8&   -1.31&  0.11&   0.26&    ...&   0.25&   0.25&  0.01&   0.30& -0.02& -0.28& -0.52& -0.18& -0.33&    0.00&  0.52\\
G271-162&   -2.33&   ...&   0.63&   ...&    ...&   0.33&   0.40&    ...&  0.48&  -0.44&  ...& -0.02&  ...& ...&...\\
G5-1&   -0.91& -0.03&   0.13&  0.26&   0.15&   0.15&   0.06&   0.22& -0.02&  -0.17& -0.47& -0.08& -0.27&  -0.21&   0.33\\
G123-9&   -1.16& -0.15&   0.25&  0.34&   0.32&   0.28&   0.17&   0.23& -0.04&  -0.18& -0.62& -0.12& -0.44& 0.18&   0.48\\
G16-20&   -1.59& -0.09&   0.28&   ...&   0.44&   0.39&  0.05&  0.21&  -0.11&  -0.21&  -0.38&  0.06&   ...&  0.07&   0.48\\
G204-30&   -0.83&  0.03&   0.26&  0.23&   0.18&   0.19&  0.26&  0.24&   0.00&  -0.16&  -0.41&  0.06&  -0.20& -0.03&   0.32\\
G37-26&   -1.89& -0.04&   0.24&   ...&   0.44&   0.26& -0.06&   ...&    ...&  -0.35&  -0.40& -0.26&   ...&   ...&...\\
HIP96115&   -2.66&   ...&   0.33&   ...&    ...&   0.71&  0.25&  0.48&   0.17&  -0.38&  -0.43& -0.08&   ...&  -0.42&...\\
HIP4754&   -1.72&   ...&   0.37&   ...&   0.48&   0.56&  0.08&   ...&   0.39&  -0.14&  -0.39&  0.15&  -0.55&  -0.18&   0.40\\
\hline
\end{tabular}
\end{table*}

\begin{table*}
\caption{Iron abundances and abundance ratios [X/Fe] for 14 elements for the Halo stars. }
\begin{tabular}{rrcrcrrrrrrrrrrr}
\hline \hline
Star&       [Fe\,{\sc i}/H]&    Na&    Mg&   Al&   Si&  Ca&  Sc&    Ti&    V  &  Cr  &    Mn&    Ni&   Cu&   Ba    &  Eu \\
\hline
 HIP 3026&   -1.20&  0.00&  0.10& 0.28& 0.26& 0.36& 0.19& 0.19&  0.09& -0.22& -0.52& -0.12& -0.55&   0.33& 0.50\\
 HIP 3430&   -1.81&   ...&  0.07&  ...&  ...& 0.27&-0.10& 0.13& -0.09& -0.24&   ...& -0.34&   ...&  -0.36& ...\\
 HIP10449&   -0.89&  0.00&  0.31& 0.32& 0.21& 0.20& 0.17& 0.20&  0.35& -0.12& -0.41& -0.03& -0.27&   0.13&  0.19\\
 HIP15904&   -1.38&  0.12&  0.28& 0.22& 0.31& 0.25& 0.12& 0.26&  0.23& -0.09& -0.18&  0.10&   ...&  -0.01&  0.50\\
 HIP16404&   -1.98&   ...&  0.26&  ...&  ...& 0.17&-0.08& 0.31&  0.08& -0.08& -0.55& -0.21& -0.27&  -0.44&  0.54\\
 HIP38541&   -1.62& -0.15&  0.26&  ...& 0.45& 0.28& 0.02& 0.24&  0.20& -0.13& -0.63& -0.16& -0.62&  -0.10&  0.32\\
 HIP42887&   -1.29&   ...&  0.26&  ...& 0.20& 0.27& 0.01&  ...&  0.25& -0.21&  ...& -0.17&   ...&  -0.19&  0.30\\
 HIP80837&   -0.63&  0.06&  0.34& 0.22& 0.13& 0.19& 0.01& 0.24&  0.22& -0.09& -0.40& -0.02& -0.11&  -0.29&  0.03\\
 HIP86321&   -0.86&  0.00&  0.11& 0.05& 0.02& 0.16& 0.00& 0.07&  0.01& -0.17& -0.48& -0.17& -0.37&   0.21&  0.20\\
 HIP86443&   -2.21&   ...&  0.30& ...&  ...& 0.33& 0.38&  ...&  0.53& -0.05&   ...& -0.07&   ...&  -0.40&  0.73\\
 HIP94449&   -1.02&  0.07&  0.26& 0.36& 0.17& 0.31& 0.20& 0.33&  0.23&  0.03& -0.48& -0.04& -0.47&   0.18&  0.37\\
 HIP100568&   -0.96& -0.26&  0.08& 0.16& 0.07& 0.13& 0.01& 0.13& -0.10& -0.10& -0.61& -0.18& -0.72&  0.04&  0.51\\
 HIP100682&   -2.11&   ...&  ...&   ...&  ...&  ...&  ...&  ...& -0.15&   ...& -0.49&   ...&  ...& -0.64&   ...\\
 HIP100792&   -1.01& -0.07&  0.19&-0.09& 0.08& 0.18&-0.05& 0.18&  0.05& -0.12& -0.49& -0.03& -0.50& -0.01&  0.25\\
 HIP105488&   -1.26& -0.01&  0.31& 0.13& 0.28& 0.24& 0.17& 0.25&  0.12& -0.10& -0.51&  0.00&   ...& -0.10&  0.49\\
 HIP106924&   -1.81&  0.10&  0.33& 0.59& 0.13& 0.26& 0.30& 0.35&  0.14& -0.12& -0.63& -0.14&   ...&  0.11&  0.36\\
 HIP111195&   -1.63&  0.10&  0.33&  ...& 0.48& 0.18& 0.30& 0.28&  0.24& -0.30&   ...&  0.01& -0.17&  0.33&  0.65\\
 HIP117041&   -0.93&  0.16&  0.34& 0.52& 0.30& 0.34& 0.10& 0.23& -0.02& -0.05& -0.35&  0.04& -0.01& -0.16&  0.09\\
\hline
\end{tabular}
\end{table*}

\subsection{Atmospheric Parameters}

Atmospheric parameters of the programme stars were
derived using  procedures described previously (Reddy et al. 2003, 2006).

\subsubsection{Effective Temperature}

The $T_{\rm eff}$ for a  star was  derived from
its $(V-K_{s})$ colour and  calibrations based
on the infrared flux method (IRFM) from Alonso et al. (1996). 
The magnitudes for $K_{s}$ were taken from  the 2MASS 
Catalog\footnote{This publication makes use of data products from the Two Micron All
    Sky Survey, which is a joint project of the University of Massachusetts
    and the Infrared Processing and Analysis Center/California Institute of
    Technology, funded by the National Aeronautics and Space Administration
    and the National Science Foundation} (Cutri et al. 2003) and the $V$ magnitudes
 were taken either from $Hipparcos$ or from the  
Arifyanto et al.  and Schuster et al. catalogues.
 The 2MASS catalog provides
the magnitude $K_{s}$ ($s$ for small) not the usual $K$ magnitude which
was used in Alonso's calibrations. Differences between the two magnitudes
and the resulting effect on the final $T_{\rm eff}$ 
are very small, as shown by 
Reddy et al. (2006). In deriving $T_{\rm eff}$, reddening was taken into account.
Reddening values, E(B-V), were taken from Carney et al. (1994)
and were converted to E(V-Ks) using R = 3.1. Typical reddening for the
sample is  E(B-V)= 0.02 and in very few
cases E(B-V) is as large as 0.05. An  E(B-V) = 0.05  corresponds to a
200~K correction.

Comparison of the $T_{\rm eff}$ derived from  photometry and
from spectroscopic criteria shows that
the mean difference ($\Delta T_{\rm eff}$) is just 1.2 K with $\sigma$ = 158 K. 
Five stars  show a much larger
difference ($\geq$ 200 K). By excluding the five stars, 
 $\Delta T_{\rm eff}$ is $\approx$ $-$29 k with $\sigma$ = 88 K i.e., 
the photometric $T_{\rm eff}$ is cooler on average by 29 K than
the spectroscopic $T_{\rm eff}$.
 We have adopted the photometric $T_{\rm eff}$ for most of the stars.
For the three stars (G62-52, HD 94449, HIP 117041), 
for which $\Delta T_{\rm eff}$  $\geq$ 200 K, 
we used the spectroscopic $T_{\rm eff}$. Adopted $T_{\rm eff}$'s are given in Tables~1 and 2.  

\subsubsection{Surface gravity and Ages}

Adopted surface gravities (log $g$) are derived using the standard relation
between log $g$, mass,  $T_{\rm eff}$ and distance (Reddy et al. 2006). Distances for all the
stars are provided either from the $Hipparcos$ catalogue or from the source catalogues. Stellar mass
has been estimated using the combination of $T_{\rm eff}$, M$_{\rm v}$ and the stellar theoretical
isochrones. For this study, Yonsei-Yale stellar tracks (Demarque et al. 2004) are used.
Tracks are computed for a given metallicity and stellar mass (in steps of 0.01) for 
the constant $\alpha$-enhancement of
[$\alpha$/Fe]=0.2 for all the stars with [Fe/H] $\leq$ $-$1.0. Appropriate stellar masses and ages are found
by evaluating how well the tracks coincide with the star's given $T_{\rm eff}$ and M$_{\rm v}$. 
Derived log $g$ values and the ages are given in Tables~1 and 2. For many of the programme stars
ages were not derived either due to a star's position closeness to 
the main sequence 
or near the hook
in the tracks.
Errors in the ages
are estimated using the uncertainties in $T_{\rm eff}$, metallicities, and parallaxes.

The log $g$ values derived from parallaxes are compared with the gravities derived using 
the standard spectroscopic method of  matching the abundances from
neutral and singly ionized atoms of iron. 
%The Tables list both log $g$
%determinations. 
In typical cases, the two methods yield consistent results. Where differences are large, the
spectroscopic analysis is likely in error because the sample of Fe\,{\sc ii} lines is
small and limited to very weak lines. 
In some cases, we have just 2-3 Fe\,{\sc ii} lines
with an W$_{\rm \lambda}$ $\approx$ 3 m\AA. In some cases,  the
`parallax' log $g$ is very uncertain, and then spectroscopic log $g$
is adopted if based on 3 or more Fe\,{\sc ii} lines of moderate strength.
The adopted gravities in the abundance analysis are given in Table~1 and 2. The log $g$ values
adopted from spectroscopic analysis are marked with symbol `s'.

\subsubsection{Microturbulence}

The microturbulence $\xi_{t}$
is estimated using neutral lines of either Fe\,{\sc i} or Ni\,{\sc i} or 
both which
have a relatively large number of lines with a range in $W_{\lambda}$s. The value
of $\xi_{t}$ for a given star is the value for which abundances of different lines
for a given atom
are independent of  W$_{\lambda}$.
 The measured $\xi_{t}$ values range 0.80 - 1.25 km s$^{-1}$.
 For very metal-poor stars, there are
 few lines and/or an inadequate range in $W_{\lambda}$ 
to measure $\xi_{t}$. For those stars, we have assumed  $\xi_{t}$ = 1 km s$^{-1}$.

\begin{figure}
\includegraphics[width=8cm,height=8cm]{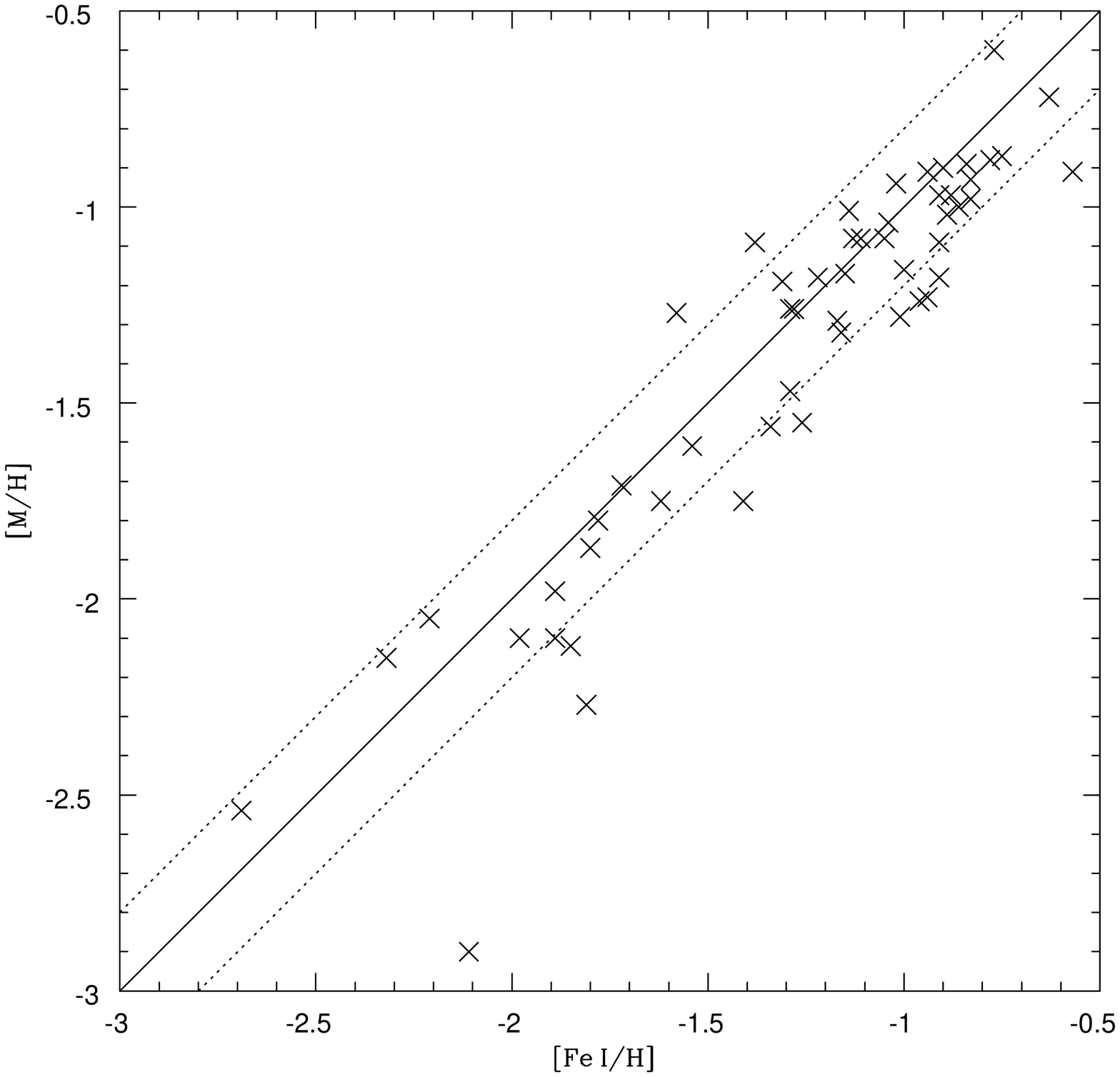}
%\epsffile{mwtd_schprob.ps}
%\epsfxsize=18truecm
\caption{ Comparison of photometric ([M/H]) and 
spectroscopic ([Fe\,{\sc i}]) metallicities. Dotted lines 
represent 0.1~dex in metallicity on either side of the solid line.}

\end{figure}

\subsubsection{Metallicity}

The  Fe abundance is
derived from Fe\,{\sc i} transitions. This follows the procedure adopted
earlier (Reddy et al. 2003, 2006).
[Fe\,{\sc i}/H] was used as metallicity.
Metallicities [M/H] are
derived using Str\"{o}mgren photometry ($uvby$) and Alonso et al.'s (1996)
calibrations. The [M/H] values are compared with the spectroscopic
[Fe\,{\sc i}/H] (Figure~8) assuming a solar abundance $\log \epsilon$(Fe)$_\odot$ = 7.45.
The [Fe\,{\sc i}/H] values are
systematically  more metal-rich than [M/H] values with a 
mean difference of 0.10$\pm$0.20. Two halo stars (HIP 3430 and HIP 100682)
show large difference in metallicities. 

\begin{figure*}
\includegraphics[width=12cm]{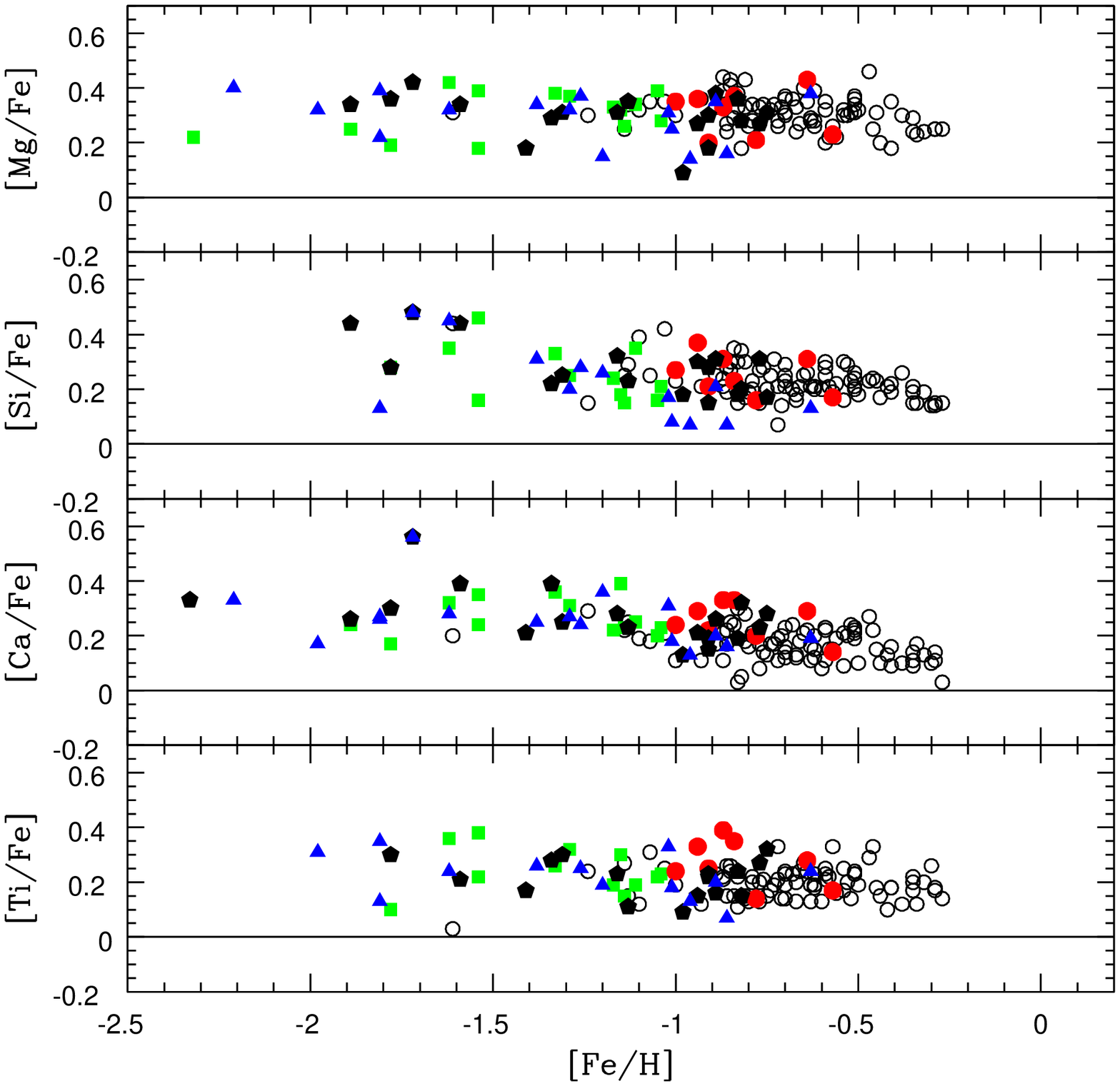}
%\epsffile{mwtd_alpha.ps}
%\epsfxsize=18truecm
\caption{ Abundance ratios   [$\alpha$/Fe] for $\alpha$ = Mg, Si, Ca, and
Ti against [Fe/H] of metal-weak thick disk  stars (green squares), 
thick disk stars (red circles), hybrid stars (black pentagons), 
halo stars (blue triangles) and  thick disc stars
from Reddy et al. (black open circles)
are shown.
}
\end{figure*}

\begin{figure*}
\includegraphics[width=12cm]{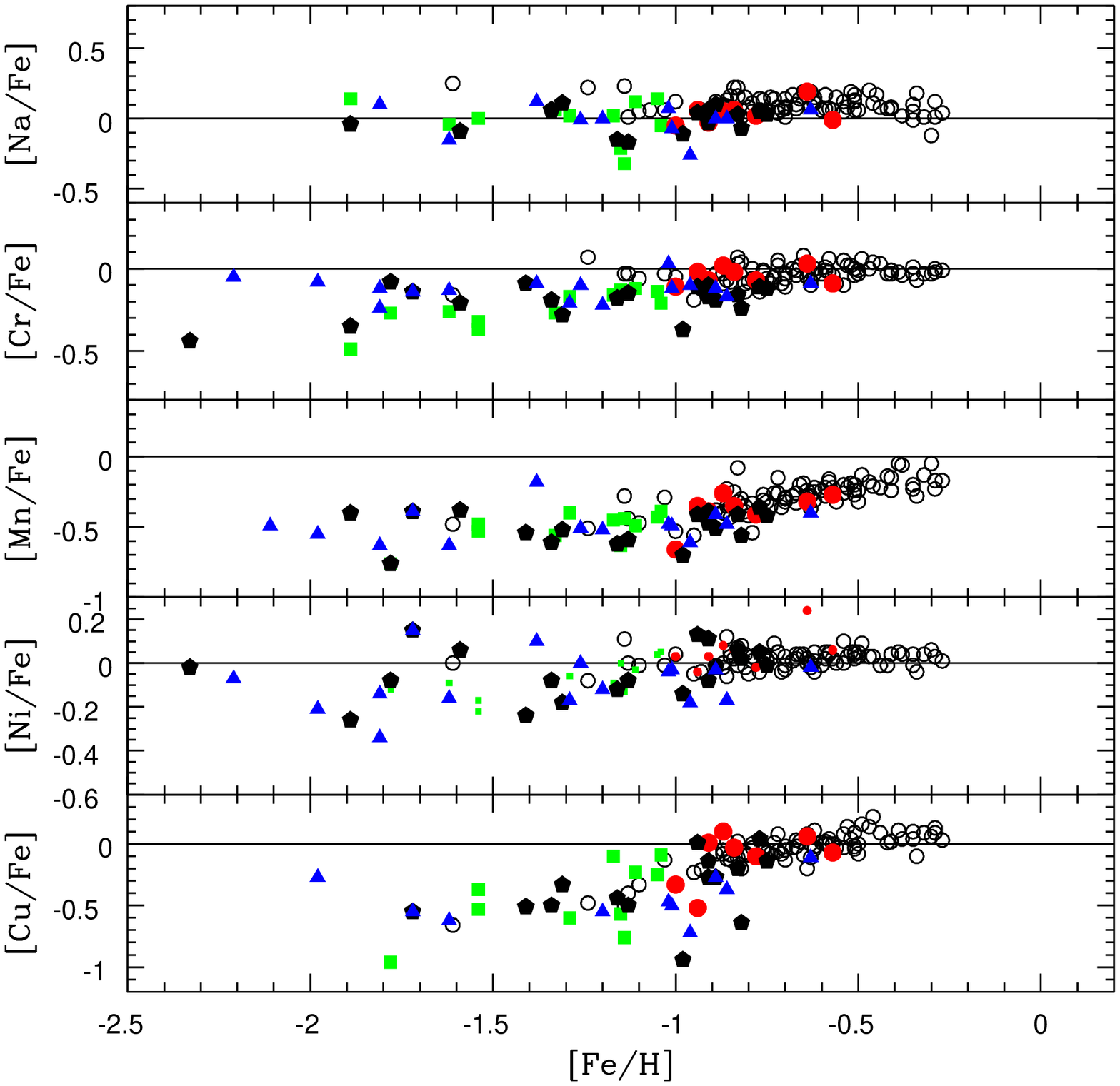}
%\epsffile{mwtd_felike.ps}
%\epsfxsize=18truecm
\caption{ Abundance ratios [X/Fe] against [Fe/H] for X = Na, Cr, Mn, Ni, and Cu
 of metal-weak thick disk candidates  (green squares), 
thick disk stars (red circles), hybrid stars (black pentagons), 
halo stars (blue triangles) and  thick disc stars
from Reddy et al. (black open circles)
are shown.
}

\end{figure*}

\subsection{Elemental Abundances} 

For the abundance determinations, we have retained and modified the line list used  earlier (Reddy et al. 2003, 2006).
This line list
was chosen such that lines are useful
for stars with [Fe/H] ranging from about $-$0.8 to $+$0.2. 
Stars in the present survey are  metal-poor with
[Fe/H] as low as  $-$2.5. Some of the lines from our original list
are very weak or absent at the metal-poor end of the sample. 
This necessitated a search for other lines
so that an element throughout the [Fe/H] range of
the sample is represented by the same lines.
The $gf$-values for the new lines are taken from
R.E. Luck's compilation (private
communication) and adjusted slightly so that the new and original line
list gave the same abundances for a few stars with [Fe/H] $\approx$ $-$1.0 whose spectra
contain all the lines. In addition to $gf$ values,
hyperfine splitting ($hfs$)  must be considered for some elements. 
Lines of elements Sc, V, Mn and Cu
are affected by $hfs$. For Mn and Cu, we have used the same lines as in Reddy et al. (2003)
but for Sc and V three new lines each are added: 5031.0\AA, 5526.8\AA, 
5657.9\AA\ lines for Sc\,{\sc ii}
and 4379.2\AA, 4384.7\AA\ and 4389.9\AA\ lines for V\,{\sc ii}.
The $hfs$ data for all the lines
were taken from Kurucz (1998), and analyzed the same way as in our earlier study. 
The difference
in abundances
for new and old lines is small ($\pm$0.05) which is within the errors resulting
from uncertainties in model parametesr and line strength measurements.

Abundances are given in Tables~3 and 4. The uncertainties are estimated
to be $\Delta$ [Fe/H] $\leq$ 0.10~dex and 
$\Delta$ [X/Fe] = 0.05 - 0.10 for most of the elements. The estimated
uncertainties are based on estimated errors in our derived model parameters: 
$\Delta T_{\rm eff}$ = $\pm$ 100 K,
$\Delta \rm log\, g$ = $\pm$ 0.25 cm s$^{-2}$,  $\Delta [M/H]$ = $\pm$ 0.25 dex, 
$\Delta \xi_{t}$ = $\pm$ 0.25 km s$^{-1}$, and $\delta W_{\lambda}$ = $\pm$ 2 m\AA.

\subsection{Comparisons with published studies}

Some stars in our sample have been included in
recent abundance analyses.
Three published abundance analyses are considered here:
Gratton et al. (2003a,b), Jonsell et al. (2005), and
Zhang \& Zhao (2005).
Inter-comparison of results is not only a check for
systematic errors but opens up the possibility that
samples may be combined to extend the dataset with
which to search for abundance signatures of metal-weak thick disk
candidates that may set them apart from halo stars.

Our [Fe\,{\sc i}/H] for the stars in common agree well
with other studies. For example, our values based on 15 common stars
are  metal-rich by 0.07$\pm$0.10~dex compared to Gratton et al. (2003a) values.
The mean difference between us and Jonsell et al. (2005) is 0.13$\pm$0.14 while
it is 0.04$\pm$0.22 between us and Zhang and Zhao (2005) - see Table 5.

Gratton et al. analysed 150 main sequence and turn-off stars
in a uniform manner starting from their own newly measured or published
equivalent widths from Fulbright (2000), Nissen \& Schuster (1997),
and Prochaska et al. (2000). There are 15 stars in common between us
and Gratton et al. with [Fe/H] spanning the range
from about $-$0.8 to $-$2.3. Gratton et al.'s
abundances come from a LTE analysis using Kurucz model atmospheres
with the effective temperature obtained from colours (B-V and $b-y$) with
supplementary estimates from either the H$\alpha$ profile or the
infrared flux method, the surface gravity from the absolute
magnitudes and theoretical isochrones and checked against ionization
equilibrium for neutral and singly-ionized iron. Abundances are given
with respect to a solar abundance analysis using 
the Kurucz solar model.  
Differences in $T_{\rm eff}$, $\log g$, [Fe/H], and [X/Fe] for
several elements are shown in Table 5. In most cases these are satisfactorily
small and certainly comparable to the estimated
uncertainties of both analyses. However, our [Mg/Fe] values are smaller by 0.14~dex
compared to Gratton et al. Most of this can be attributed to the adopted solar Mg abundance
in the analysis. We have adopted log $\epsilon$ (Mg) = 7.64 (Reddy et al. 2003) for the Sun
and Gratton et al. used log $\epsilon$ (Mg) = 7.43 from their solar abundance analysis. 
The other contributor possibly be due to 
adoption of different sets of Mg\,{\sc i} lines in the analysis. Gratton et al.
used 10 Mg\,{\sc i} lines against four in our study.

Jonsell et al. derived stellar parameters for 43 metal-poor
stars from Str\"{o}mgren
photometry, and stellar isochrones using the {\it Hipparcos}
parallax. {\it MARCS} models were used in an LTE analysis of lines
also used to analyse the solar spectrum. Five  of our stars 
were analysed by Jonsell et al. with the quintet spanning the
[Fe/H] range from about $-$1.0 to $-$1.6. 
Differences in $T_{\rm eff}$, $\log g$, [Fe/H], and [X/Fe] for
several elements are satisfactorily
small in all cases and certainly comparable to the estimated
uncertainties of both analyses (Table~5). However, our Fe and Ba ratios
are significantly larger compared to Jonsell et.al.  
This difference in Fe may be attributed to use of
different sets of Fe\,{\sc i} lines in the studies. 
We have used 43 well defined Fe\,{\sc i} lines 
against 18 lines
in Jonsell et al. Only 4 lines are common with us.
Another reason could be the difference of 0.05~dex in 
the adopted solar abundances. Jonsell et al. have adopted log $\epsilon$ (Fe) = 7.50 for the Sun.
On the other hand the large difference in Ba ratio is not understood. However, Jonsell et al.
ratio is smaller by same magnitude compared to other three recent studies 
(see their Table~6 and references
therein).

Zhang $\&$ Zhao analysed 31 metal-poor stars obtaining effective
temperatures from colours (B-V, V-K, $b-y$), surface gravities from
stellar isochrones and {\it Hipparcos} parallaxes. Stellar LTE abundances
were referenced to Grevesse $\&$ Sauval's (1998) solar abundances in order
to obtain the differential abundances [X/H] except that for iron solar
abundance of 7.51 was adopted. For the 11 stars in common with our
sample, Table 5 gives the comparisons. In the case of [Na/Fe], we give
the result using Zhang $\&$ Zhao's LTE Na abundance. Our results agree well with
the results from Zhang $\&$ Zhao except for two elements Mn and Ni. Exact cause for
such discrepancy is not known but it may be attributed to different line sets used, and the adopted
sources of $hfs$
structure and atomic data.

Jonsell et al. (their Table 6)
provide a comparison of abundances obtained for common stars  from
multiple published works showing that for  many elements differences
between LTE analyses are 0.10 dex or better. Our comparisons in
Table~5 echo
this level of consistency in most cases.  

\begin{table}
\centering
\caption{ Mean differences and standard deviations of the
abundance ratios [X/Fe] for 15 stars that are common with
Gratton et al. (2003a), 5 stars with Jonsell et al. (2005), and 
11 stars with Zhang $\&$ Zhao (2005)}

\begin{tabular}{@{}lrrrrrr@{}}
\hline \hline
         &\multicolumn{2}{c}{Ours - Gratton} &\multicolumn{2}{c}{Ours-Jonsell} & 
\multicolumn{2}{c}{Ours-Zhang}  \\
Quantity & diff.  & $\sigma$ &  diff.    & $\sigma$ & diff.  & $\sigma$ \\
\hline
T$_{\rm eff} $  & -26     & 99     &  39      & 33    & 12   & 79   \\
log $g$         & -0.03   & 0.15   &0.03      & 0.20  & 0.25 & 0.25 \\
${\rm [Fe/H]}$  &  0.07   & 0.10   &0.14      & 0.07  & 0.04 & 0.22  \\
${\rm[Na/Fe]}$  & -0.01  & 0.12    &  0.08    & 0.12  & 0.12 & 0.22   \\
${\rm[Mg/Fe]}$  & -0.14  & 0.10    & -0.08    & 0.07  &-0.04 & 0.29       \\
${\rm[Al/Fe]}$  &  ....  & ....    & -0.13    & 0.0   & .... & .... \\
${\rm[Si/Fe]}$  & -0.05  & 0.17    & -0.06    & 0.03  & 0.10 & 0.20       \\
${\rm[Ca/Fe]}$  & -0.05  & 0.04    & 0.01     & 0.08  & 0.03 & 0.13       \\
${\rm[Sc/Fe]}$  &  0.09  &0.15     & -0.02    & 0.05  & 0.05 & 0.23     \\
${\rm[Ti/Fe]}$  & -0.01  & 0.11    & ....     & ...   & -0.04 &0.08     \\
${\rm[V/Fe]}$   &  0.09  &0.15     & ...      & ...   & -0.21 & 0.19     \\
${\rm[Cr/Fe]}$  & ....  & ....    & ...      & ...   & -0.11  & 0.07    \\
${\rm[Mn/Fe]}$  & -0.10  &0.15     & ...      & ...   &-0.29  & 0.15      \\
${\rm[Ni/Fe]}$  & -0.05  & 0.11    & 0.07     & 0.07  & -0.22 & 0.18     \\
${\rm[Ba/Fe]}$  & ....   & ....    & 0.26     &  0.03 & 0.01  & 0.20       \\
\hline

\end{tabular}
\end{table}

\begin{table}
\centering
\caption{ Mean abundance ratios [X/Fe] and the standard deviations ($\sigma$)
of the measured elements for the 18 halo, 14 metal-weak thick disk stars (Table~2; this study) and 
66 thick disc stars (Reddy et al. 2006).} 

\begin{tabular}{@{}lrrrr@{}}
\hline \hline
              &Thick disc &metal-weak thick disk   &  Halo \\
\hline
${\rm[Na/Fe]}$  & 0.11$\pm$0.06  &  $-$0.01$\pm$0.10    &   $-$0.01$\pm$0.15   \\ 
${\rm[Mg/Fe]}$  & 0.31$\pm$0.06    & 0.32$\pm$0.07     & 0.31$\pm$0.09  \\ 
${\rm[Al/Fe]}$  & 0.27$\pm$0.08     &  0.30$\pm$0.11    & 0.21$\pm$0.10    \\ 
${\rm[Si/Fe]}$  & 0.23$\pm$0.06     &  0.27$\pm$0.09    & 0.23$\pm$0.15   \\ 
${\rm[Ca/Fe]}$  & 0.18$\pm$0.06    & 0.27$\pm$0.06     & 0.25$\pm$0.07    \\ 
${\rm[Ti/Fe]}$  & 0.20$\pm$0.06    & 0.25$\pm$0.08     & 0.23$\pm$0.08    \\ 
${\rm[V/Fe]}$   & 0.11$\pm$0.08    & 0.07$\pm$0.10      & 0.12$\pm$0.16   \\ 
${\rm[Cr/Fe]}$  & 0.03$\pm$0.05    &$-$0.21$\pm$0.12     &$-$0.13$\pm$0.08     \\ 
${\rm[Ni/Fe]}$  & 0.02$\pm$0.04    &$-$0.08$\pm$0.09    & $-$0.08$\pm$0.11    \\ 
${\rm[Eu/Fe]}$  &  0.38$\pm$0.13    &  0.44$\pm$0.24    & 0.35$\pm$0.15    \\  
\hline

\end{tabular}
\end{table}

\begin{figure}
\includegraphics[width=8cm]{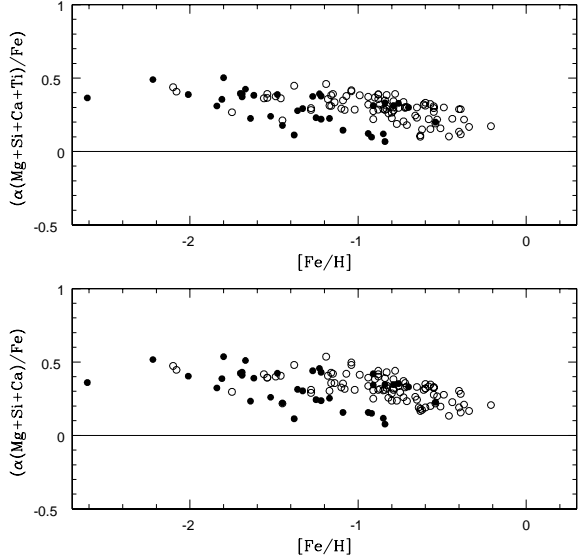}
%\epsffile{mwtd_heavy.ps}
%\epsfxsize=18truecm
\caption{ The run of [$\alpha$/Fe] versus [Fe/H] from Gratton et al. (2003a, b)
for their dissipative (open circles) and accretion (filled circles) components.
The [$\alpha$/Fe] are computed as the mean of [Mg/Fe], [Si/Fe], [Ca/Fe] and [Ti/Fe]
for the top panel and [Mg/Fe], [Si/Fe] and [Ca/Fe] for the bottom panel.}
\end{figure}

\begin{figure}
\includegraphics[width=8cm]{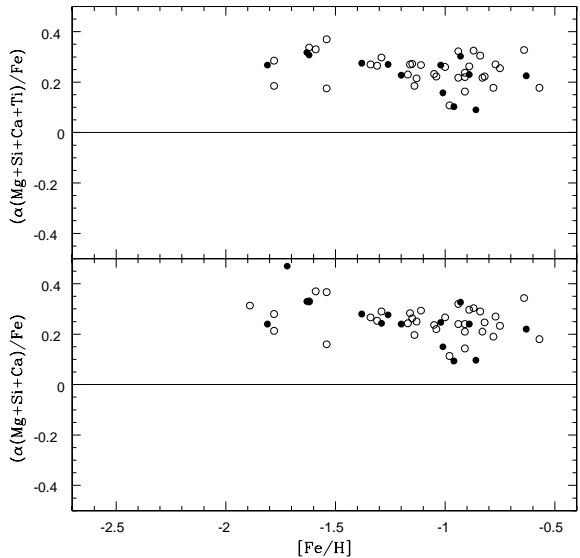}
%\epsffile{mwtd_heavy.ps}
%\epsfxsize=18truecm
\caption{ The run of [$\alpha$/Fe] versus [Fe/H] from this study.
for the dissipative (open circles) and accretion (filled circles) components.
The [$\alpha$/Fe] are computed as the mean of [Mg/Fe], [Si/Fe], [Ca/Fe] and [Ti/Fe]
for the top panel and [Mg/Fe], [Si/Fe] and [Ca/Fe] for the bottom panel.}
\end{figure}

\begin{figure}
\includegraphics[width=8cm]{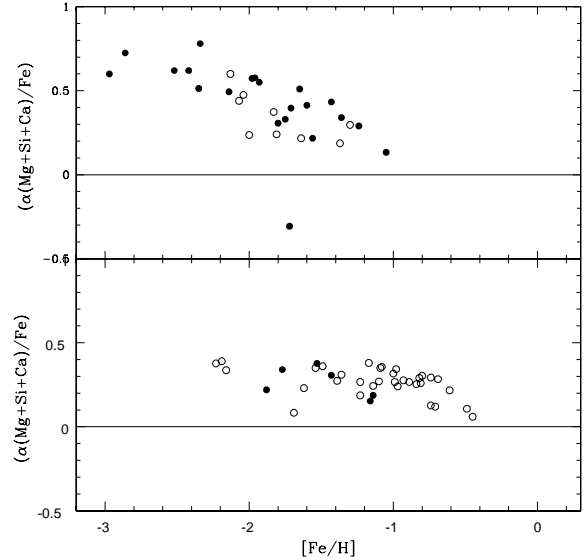}
%\epsffile{mwtd_heavy.ps}
%\epsfxsize=18truecm
\caption{ The run of [$\alpha$/Fe] versus [Fe/H] from Zhang $\&$ Zhao (2005) (top panel)
and from Jonsell et al. (bottom panel) 
for the dissipative (open circles) and accretion (filled circles) components.
The [$\alpha$/Fe] are computed as the mean of [Mg/Fe], [Si/Fe], [Ca/Fe]. 
In both the studies Ti abundance was determined for very few stars and hence is not
considered. The anomalous low point is the $\alpha$-poor star BD+80$^{\o}$ 245.}

\end{figure}

\section{Discussion}
Dissection of the Galactic stellar population into principal
components relies on combinations of differences in
age, kinematics,   composition and location.
Differentiation of the standard thick disc
from the
thin disc is remarkably clear using all these properties: the
thick disc is older than the thin disc, the thick disc lags the
thin disc by about 50 km s$^{-1}$, and is not only slightly
metal-poor relative to the thin disc (say, mean [Fe/H] $\simeq -0.6$
to $\simeq 0$) but there are striking differences in relative
abundances ([X/Fe]), and the scale height of the thick disc
is larger than that of the thin disc. A similar description of contrasts may
be given to differentiate the halo from the thin disc. 
 Isolation of the metal-weak thick disk and a
determination of its status either as  the metal-poor extension of the thick disc,
a collection of halo stars, or 
as a discrete Galactic component (or mixture of components) must use the same
quartet of characteristics. 

  Unlike differentiation of the thick
from the thin disc where differences in the UVW velocities
and their dispersions allow a statistical  separation  of thin and thick disc stars,
the kinematical differences between the metal-weak thick disk and thick
disc on the one hand and the metal-weak thick disk and halo on the other hand are not
transparently  clear at present. If one assumes that there are no
kinematical differences between thick disc and a metal-weak thick disk, Figure~2c shows that metal-weak thick disk
stars are rare in the solar neighbourhood relative to thick
disc stars, if selection effects are set aside.
 Alternatively, if either the $V_{\rm rot}$ of the metal-weak thick disk is
similar to that of Schuster et al.'s  second group of thick disc stars
or the $V_{\rm rot}$ of the metal-weak thick disk decreases with [Fe/H],  metal-weak thick disk stars
share kinematics with halo stars increasing the difficulty in isolating the metal-weak thick disk
star from a halo star. 
Our criteria for $|U_{\rm LSR}|$, V$_{\rm rot}$, and $|W_{\rm LSR}|$ are
an attempt to enhance selection of  metal-weak thick disk candidates over halo stars  from
combined catalogue.

Age differences between halo and metal-weak thick disk stars will be difficult to
establish. It is known that thick disc stars are older systematically
than thin disc stars with thick disc and halo star ages approaching the
WMAP age of the Universe (Schuster et al. 2006); the age
difference between the two populations does not exceed the uncertainty
of the age determinations.  Our age estimates (Tables 1, 2) for metal-weak thick disk 
candidates also in the mean approximate the WMAP age and, thus,
halo, thick and metal-weak thick disk candidate stars cannot be set apart by
age using estimates of the present precision.

Perhaps, a chance of establishing that metal-weak thick disk stars are
a distinct population may rest with the relative abundances
[X/Fe]. However, comparison of the abundances for our
halo and metal-weak thick disk candidates shows no clear differences in
[X/Fe]. This is summarized by Table~6 giving  the
mean and standard deviations for [X/Fe]  for both samples. There is
not a single significant difference. (Note that several potentially
interesting elements were not measured in our study: C, N, and O, in
particular.)  Both samples span  similar ranges in [Fe/H] and
atmospheric parameters so that non-LTE effects are very unlikely
to have exerted a differential effect on the halo-metal-weak thick disk comparison.
This close similarity in composition allows of two possibilities,
either the metal-weak thick disk stars are halo stars under a 
redundant label or the metal-weak thick disk and halo stars are  distinct populations
with indistinguishable compositions at a given [Fe/H]. This is quite a contrast from the recent
study, yet to be published, by Nissen $\&$ Schuster (2008). They have showed a clear 
[X/Fe] abundance difference among
stars with space velocities very similar to halo and thick disk:
one group of stars shows high [$\alpha$/Fe] and the other shows 
significantly lower [$\alpha$/Fe] values. They claim that this may be due to two different 
population of stars within halo originating from different sources.

If the metal-weak thick disk is the metal-poor tail of the thick disc, 
a necessary condition would seem to be that
the compositions provide a continuous trend in [X/Fe] versus
[Fe/H]. This is the case, as shown for several elements 
in Figures~8 and 9. For the metal-weak thick disk candidates and thick disc stars, abundances merge
smoothly across [Fe/H] $\simeq -1$ to within about $\pm$0.1 dex. 
The mean [X/Fe] for the 66 thick disc stars 
with $-1.0 < $[Fe/H]$ < -0.5$ from Reddy et al.(2006)  show
good agreement with the candidate metal-weak thick disk stars from Table~1.
Unfortunately, this result is insufficient to establish the metal-weak thick disk
as the tail of the thick disc population because, as just noted,
the metal-weak thick disk and halo share the same [X/Fe] results. 

Gratton et al (2003b) proposed a differentiation by kinematics
of those stars that do not belong to the thin disc.
 Galactic components - dissipative and accretion - were
introduced with the definitions:

dissipative: $V_{\rm rot} > 40$ km s$^{-1}$ and $R_{\rm max} < 15$ kpc
where $R_{\rm max}$ is the maximum distance of star from the Galactic
centre in its computed orbit.

accretion: $V_{\rm rot} < 40$ km s$^{-1}$

The definitions dissipative and accretion  were intended to describe the
mode of formation of the component.  Gratton et al. suggested that
there was a  systematic difference in the run of [$\alpha$/Fe]
versus [Fe/H] between the components - Figure~10 reconstructs
their illustration from their data where we give two
choices for the $\alpha$ index. The selection of 40 km s$^{-1}$ as the 
boundary between the dissipative and accretion components was
apparently an arbitrary one. 
The dissipative component seems to show a rather tight run of [$\alpha$/Fe]
with [Fe/H] for [Fe/H] $< -1$, and
the accretion component to show a greater dispersion with the upper boundary 
coinciding with the dissipative results. For [Fe/H $> -1.2$, 
an accretion-dissipative difference was shown first by
Nissen \& Schuster (1997). The difference was effectively noted
by Fulbright (2000) who claimed a correlation
between the Galactic rest frame velocity 
($v_{RF} = [U_{LSR}^2+(V_{LSR}+220)^2+W_{LSR}^2]^{1/2}$) and  [X/Fe]
for many elements X including the $\alpha$ elements. The mean [$\alpha$/Fe]
value dropped by about 0.1  dex for
stars with $v_{RF} > 300$ km s$^{-1}$ relative to mean values for stars
 of lower  $v_{RF}$. 

Since the thick disc stars define the high [Fe/H] end of the
dissipative results, this difference offers a tool with which to
investigate the population to which the metal-weak thick disk candidates belong.
However, Figure~11 shows that stars in our samples show the same
run of [$\alpha$/Fe] versus [Fe/H] for dissipative and accretion
components. 
Zhang \& Zhao's (2005) sample (Figure~12) suggests the reverse conclusion
about the two components: the accretion component may have higher
[$\alpha$/Fe] than the dissipative component. Jonsell et al.'s sample (Figure~12) for
[Fe/H] $< -1$ shows no difference between the two components. In short,
the distinction between accretion and dissipative components 
is not crisply defined as
suggested by Gratton et al. The lack of a clear difference
among these components  precludes assignment of metal-weak thick disk
stars to a particular component. 

\begin{table}
\centering
\caption{ Mean orbital parameters for R$_{\rm m}$, e and Z$_{\rm max}$ for the
stellar samples in Tables 1 and 2. Number of stars ($n$) in each group is given in
coloum 2. 
Sample of 66 thick disk$^{*}$ stars are from Reddy et al. (2006)} 

\begin{tabular}{@{}llrrr@{}}
\hline \hline
Sample          & $n$ & $< R_{\rm m} > $  & $< e > $  &  $ < Z_{\rm max} > $\\
\hline
Metal-weak thick disk           &  14         &  6.38$\pm$0.44    & 0.49$\pm$0.10 & 0.59$\pm$0.40  \\ 
Thick disc (this study)     &  8          &  6.45$\pm$0.40    & 0.45$\pm$0.07 & 0.81$\pm$0.48  \\ 
Thick disc$^{*}$ & 66          &  6.44$\pm$0.70    & 0.47$\pm$0.10 & 0.65$\pm$0.27  \\ 
Hybrid         &  20         &  6.26$\pm$1.50    & 0.65$\pm$0.11 & 0.55$\pm$0.53  \\
Halo           &  18         &  6.59$\pm$2.60     & 0.87$\pm$0.16 & 0.65$\pm$0.35  \\
\hline

\end{tabular}
\end{table}

An alternative to the dissipative/accretion components is provided by 
characterization of the computed orbits of the stars. 
Key parameters
are the mean of the apogalactic and perigalactic distances ($R_{\rm m}$), the
eccentricity (e) and the maximum height from the Galactic plane ($Z_{\rm max}$).
These values are given in Tables 1 and 2. Table~7 gives the mean values of $R_{\rm m}$,
e, and
$Z_{\rm max}$ for the stars in Tables 1 and 2 
and for 66 thick disc stars in Reddy et al. (2006).

There appears to be an interesting message to be drawn from Table~7. The candidate metal-weak thick disk stars
share the mean orbital parameters with the eight thick disc stars and, more significantly
with the large sample of thick disc stars from Reddy et al. (2006). Noticeable
is the agreement between the mean eccentricity
 and its dispersion among these three groups. 
The mean eccentricity for the hybrid
 stars (Table~1) is intermediate between that of the thick disc (Table~1)
and the halo stars in Table~2. Even the hybrid stars with [Fe/H] $\geq -1.3$ 
have in the main eccentricities greater than values typical of thick disc
stars. Thus, the hybrid stars are predominantly representatives of the halo.

Examination of the orbital parameters (Table~7) suggests that the velocity criteria
that led to the collection of metal-weak thick disk candidate
 stars in Table~1 may isolate a sample to which
the metal-weak thick disk is a major contributor. The [Fe/H] for this sample runs from $-$1.05 to $-$2.32;
stars with [Fe/H] $\geq$ $-$1.0 were assigned as thick disc stars.
Eight of the 14 metal-weak thick disk stars
have [Fe/H] $\geq$ $-$1.3 and might most plausibly be seen as a mild extension of the [Fe/H]
distribution of the thick disc. The other six stars have [Fe/H] from $-1.5$ to 
$-2.3$ with no significant difference  
in the kinematical and orbital parameters from
the eight with [Fe/H] $\geq$ $-$1.3. This eight are possible representatives of the metal-weak thick disk population.

Metal-weak thick disk candidates according to our velocity criteria are present in other
studies, as indeed would be expected. Among Gratton et al.'s (2003a,b) collection
of 150 stars, we find using their LSR-velocities 14 metal-weak thick disk candidates
of which four are in our sample. Seven of the 14 have [Fe/H] $\geq -1.2$ and
may be thick disc stars in the low [Fe/H] tail. The other seven have [Fe/H]
values from $-1.3$ to $-2.5$ for a mean value of $-1.7$. Jonsell et al.'s (2005)
sample of 43 stars includes eight metal-weak thick disk candidates with five having [Fe/H]
$\geq -1.2$. The other three with [Fe/H] from $-1.4$ to $-2.2$ are metal-weak thick disk 
candidates. Zhang \& Zhao's (2005) sample provides two metal-weak thick disk candidates but both
are in Table 1. 

\section{Concluding remarks}

Our collection of 14 stars in Table 1 are offered as metal-weak thick disk
candidates but by no means as certain members of the elusive metal-weak thick disk
population. We have been unable to identify a
conclusive signature distinguishing a metal-weak thick disc
star from a halo star. In terms of age, present knowledge and
precision of age determinations do not provide a discriminant
between thick disc and halo. At a given [Fe/H], the relative
abundances [X/Fe] of metal-poor stars appear to show a dispersion
no larger than the measurement uncertainties. Relative to
the uncertainties, the halo stars and our metal-weak thick disk candidates of similar
[Fe/H] are
not distinguishable. Our sole proposed discriminant involves
a combination of the LSR-velocities.

This discriminant is imperfect because kinematics of the
thick disc and halo involve overlapping distribution
functions. The overlap is essentially complete for the
$U$ and $W$ components: the thick disc velocity distributions are wholly
contained within the broader halo distributions. The overlap for the $V$ distributions
is dependent on the Fe-dependence of the
(unknown) velocity distribution of the halo stars (see the
difference between Figures~2a and 2b) and on the extrapolation of the
thick disc velocity distribution into the regime of [Fe/H] $\leq -1$, as
discussed above. The $V_{\rm rot}$ of the metal-weak thick disk candidates is comparable to
that of Schuster et al's second group of thick disc stars. 
 While our velocity criteria are intended to
favour selection of thick disc stars, contamination of the
metal-weak thick disk candidates by halo stars remains a possibility. 
The eccentricity is no more than an intriguing discriminator.
Selection of metal-weak thick disk candidate stars by their $V_{\rm rot}$ ensures that they
have a lower eccentricity than the stars of much lower $V_{\rm rot}$
in Table 2. 
%The distribution
%functions for the LSR-velocities admit of stars with velocities
%satisfying the criteria we impose on the metal-weak thick disk candidates. 

Selection effects influencing the two catalogues certainly
play a role. This is strongly suggested, as noted above, by the rather
different distributions of metal-poor ([M/H] $< -1$) stars in
the Arifyanto et al. catalogue (Figure~2a) and the Schuster et al.
catalogue (Figure~2b). In Figure~2a, the stars are rather
symmetrically distributed about $V_{\rm rot}$ = 0, the boundary
between stars on prograde and retrograde orbits. In contrast, Figure~2b
shows a preponderance of retrograde orbits and a mean $V_{\rm rot}$
decreasing with decreasing [M/H]. Setting aside the influence of
selection effects, very different models of the halo $V_{\rm rot}$
distribution versus [M/H] result from the two catalogues. In turn, these
different models would result in different conclusions about the
extension of the thick disc to [M/H] $< -1$ that result from
subtraction of the halo distribution from the observed
distribution. Most probably, neither represents the true
halo distribution.

The small sample of metal-weak thick disk candidates have the
composition established for the thick disc and the halo at their
interface. Obviously, the case for a metal-weak thick disk population needs
refinement beginning with the isolation of  a larger sample with 
reliable metallicities assuring that stars are indeed metal-weak and
reliable kinematics from accurate distances, radial velocities, and
proper motions to establish that the stars have disc-like
motions. The tools with which to meet these goals are
on the horizon.

Inevitably, the tools will enable stars of the halo, the  thick disc and its
metal-weak tail to be examined at up to several kpc from the
Sun. Examination of the populations characteristics as a function
height from the Galactic plane and distance from the Galactic
centre will provide much needed information for digestion by
theorists seeking to unravel the history of the Galaxy. Results
are now beginning to appear from analyses of the SDSS database - see, for example,
Allende Prieto et al. (2006) and
Ivezi\'{c} et al. (2007) who use the SDSS data to analyse spectra of
large numbers of F and G dwarfs at kpc distances from the Sun.
It will now be interesting to obtain detailed abundances (i.e., [X/Fe])
for samples of these faint stars.

Two considerations strongly imply that stars different in [X/Fe] at a given
[Fe/H] should be uncovered by examining such samples. First, stars in
dwarf spheroidal galaxies (dSphs) and the Magellanic Clouds have lower [$\alpha$/Fe]
at a given [Fe/H] than the Galactic halo and thick disc stars; a
compilation by Koch et al. (2008) shows that across the range of [Fe/H] from
about $-0.5$ to $-3$ the [$\alpha$/Fe] of the dSphs  is 0.2 to 0.3~dex smaller than
for the Galactic halo and thick disc stars (see their Figure~3). 
Second, the Galaxy is known to be accreting dSphs. Therefore, accretion of
stellar systems akin to the present day dSphs is expected to add stars of
lower than typical [$\alpha$/Fe] to the halo populations. Conversely, the
seeming rarity of such stars in the general local stellar population
implies that systems like the dSphs did not play a major
role in the merger history of the halo and thick disc.

This research has been supported in part by the Robert A. Welch
Foundation of Houston, Texas. We thank anonymous referee for 
the detailed review which helped to improve the paper.

{}

\end{document}